\documentclass[prr,superscriptaddress,twocolumn,showpacs,preprintnumbers,amsmath,amssymb,fixfloat]{revtex4-2}

\usepackage{graphicx}
\usepackage{amssymb}
\usepackage{dcolumn}
\usepackage{bm}
\usepackage{bbm}
\usepackage{float}
\usepackage{siunitx}

\begin{document}
\title{Quantum polyspectra for modeling and evaluating quantum transport measurements: A unifying approach to the strong and weak measurement regime}
\author{M. Sifft}
\address{Ruhr University Bochum, Faculty of Physics and Astronomy, Experimental Physics VI (AG), Germany}
 \author{A. Kurzmann}
\address{Faculty of Physics and CENIDE, University of Duisburg-Essen, Lotharstra{\ss}e 1, 47057 Duisburg, Germany}
 \author{J. Kerski}
 \address{Faculty of Physics and CENIDE, University of Duisburg-Essen, Lotharstra{\ss}e 1, 47057 Duisburg, Germany}
 \author{R. Schott}
 \address{Lehrstuhl f\"ur Angewandte Festk\"orperphysik, Ruhr-Universit\"at Bochum, Universit\"atsstra{\ss}e 150, D-44780 Bochum, Germany}
 \author{A. Ludwig}
 \address{Lehrstuhl  f\"ur Angewandte Festk\"orperphysik, Ruhr-Universit\"at Bochum, Universit\"atsstra{\ss}e 150, D-44780 Bochum, Germany}
 \author{A. D. Wieck}
 \address{Lehrstuhl  f\"ur Angewandte Festk\"orperphysik, Ruhr-Universit\"at Bochum, Universit\"atsstra{\ss}e 150, D-44780 Bochum, Germany}
 \author{A. Lorke}
 \address{Faculty of Physics and CENIDE, University of Duisburg-Essen, Lotharstra{\ss}e 1, 47057 Duisburg, Germany}
 \author{M. Geller}
 \address{Faculty of Physics and CENIDE, University of Duisburg-Essen, Lotharstra{\ss}e 1, 47057 Duisburg, Germany}
 \author{D. H\"agele}
 \address{Ruhr University Bochum, Faculty of Physics and Astronomy, Experimental Physics VI (AG), Germany}
 
\date{\today}
\begin{abstract}
Quantum polyspectra of up to fourth order are introduced for modeling and evaluating quantum transport measurements offering a powerful alternative to methods of the traditional full counting statistics. Experimental time-traces of the occupation dynamics of a single quantum dot are evaluated via simultaneously fitting their 2nd-, 3rd-, and 4th-order spectra. The scheme recovers the same electron tunneling and spin relaxation rates as previously obtained from an analysis of the same data in terms of factorial cumulants of the full counting statistics and waiting-time distributions. Moreover, the evaluation of time-traces via quantum polyspectra is demonstrated to be feasible also in the weak measurement regime even when quantum jumps can no longer be identified from time-traces and methods related to the full counting statistics cease to be applicable. A numerical study of a double dot system shows strongly changing features in the quantum polyspectra for the transition from the weak measurement regime to the Zeno-regime where coherent tunneling dynamics is suppressed. Quantum polyspectra thus constitute a general unifying approach to the strong and weak regime of quantum measurements with possible applications in diverse fields as nano-electronics, circuit quantum electrodynamics, spin noise spectroscopy, or quantum optics. 
\end{abstract}

 \maketitle
\section{Introduction}
\label{sec:introduction}
\begin{figure}[t]
	\centering
	\includegraphics[width=7.7cm]{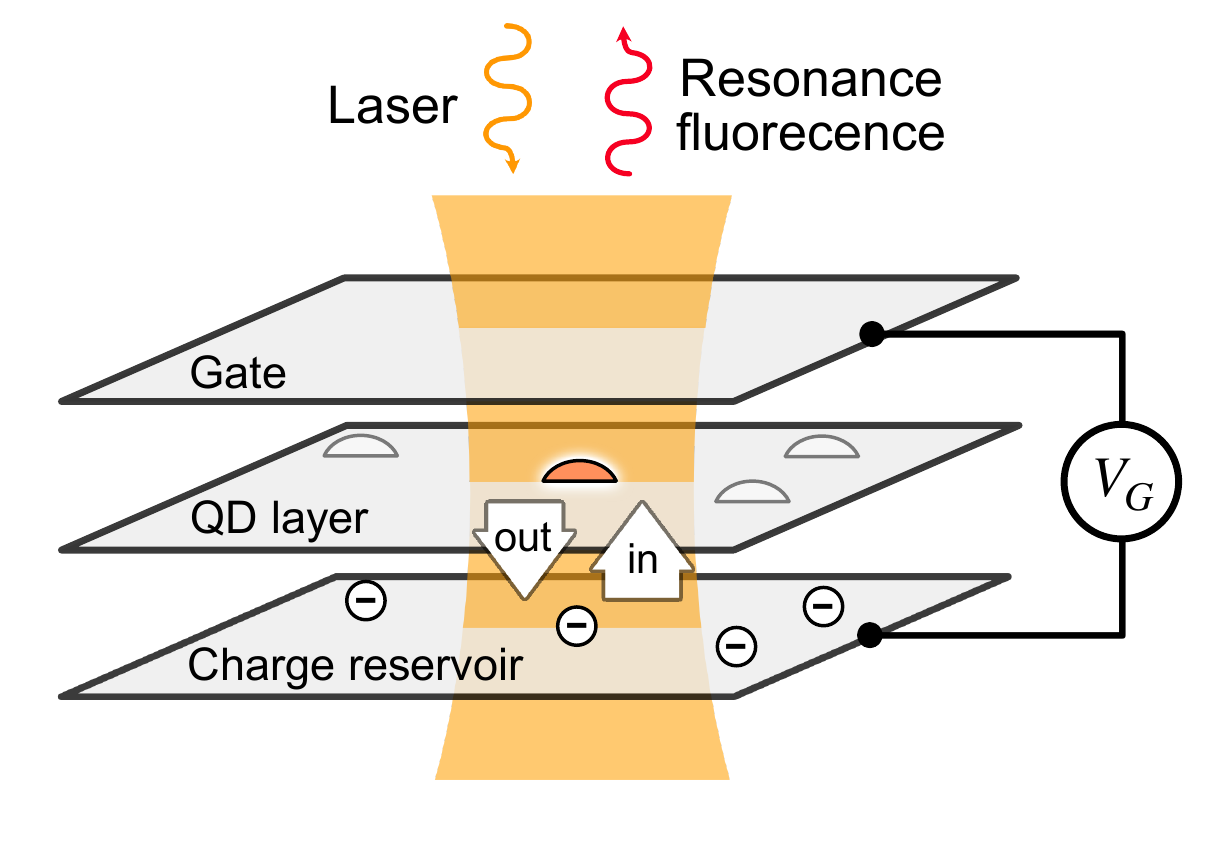}
	\caption{Tunnelung events of electrons between the semiconductor QD and the charge reservoir are monitored via the resonance fluorescence of the exciton transition.}
	\label{schema}
\end{figure}

\begin{figure}[t]
	\centering
	\includegraphics[width=6.7cm]{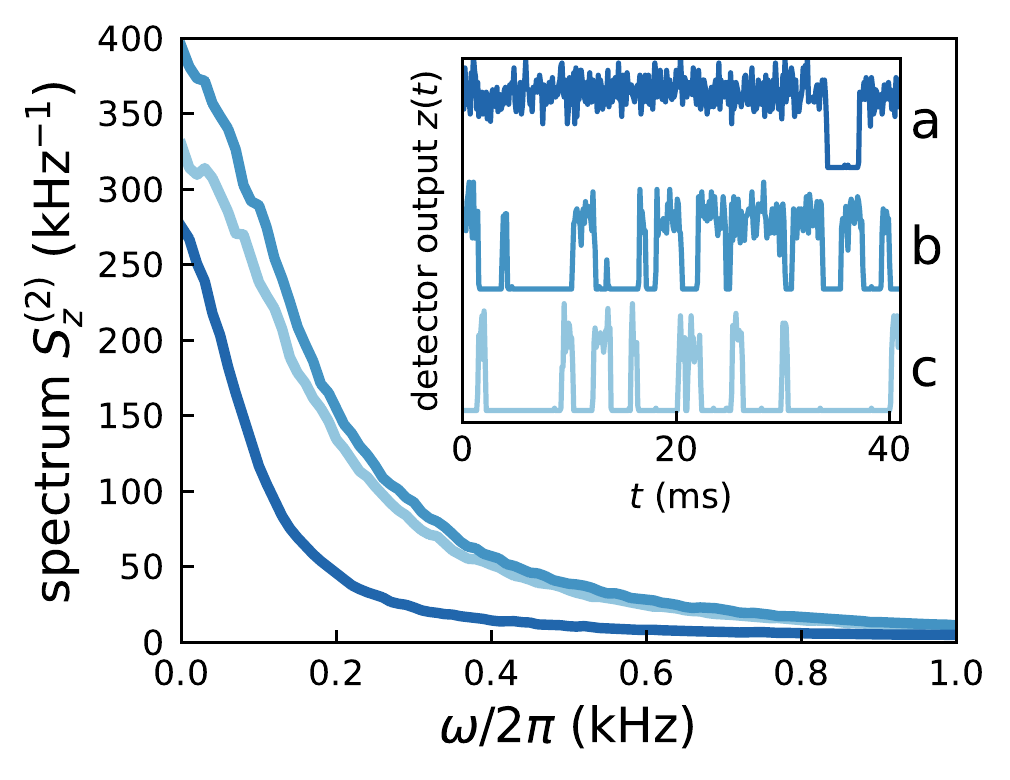}
	\caption{Power spectra $S^{(2)}_z(\omega)$ of experimental fluorescence time traces $z(t)$ (inset) for a single quantum dot at \mbox{10 T} and different gate voltages. Different regimes for tunnel rates $\gamma_{\text{in}}$ and $\gamma_{\text{out}}$ are observed: Case a: $\gamma_{\text{in}} \ll \gamma_{\text{out}}$; case b: $\gamma_{\text{in}} \gtrapprox \gamma_{\text{out}}$; and case c: $\gamma_{\text{in}} > \gamma_{\text{out}}$.}
	\label{sample_s2}
\end{figure}

Quantum measurements are at the heart of many fields in physics like quantum electronics, quantum optics, circuit quantum electrodynamics \cite{blaisNATUREPHYS2020}, or the quickly developing field of quantum sensing \cite{degenRMP2017}. In many cases, the detector output of a measurement scheme results in stochastic time-traces with information on the measured quantum system hidden in the data. Various schemes for recovering that information are employed depending on the specifics of both the quantum system and the measurement setup. In the field of quantum electronics, the dynamics of electron occupation of semiconductor quantum dots can, e.g., be measured via a so-called quantum point contact (QPC) in the vicinity of the quantum dot. The charge state of the quantum dot is immediately revealed by the strength of the probe current  \cite{ubbelohdeNATCOMM2012}. Alternatively, the occupation of an illuminated quantum dot has been measured via its resonance fluorescence \cite{kurzmannPRL2019} (see Fig. \ref{schema}). 
The resulting time traces $z(t)$ of the detector output
exhibit for both schemes telegraph noise due to quantum jumps in the occupation dynamics (see inset of Fig. \ref{sample_s2}). Jumps relating to an electron leaving the dot are then often analyzed via the so-called full counting statistics (FCS) $p(N,t)$, where $p$ is the probability that $N$ electrons have left the quantum dot in the time interval $t$ \cite{levitovJMP1996,bagretsPRB2003}. 
The counting statistic $p(n,t)$ is given for a simple tunnel-barrier and fixed $t$ by a Poisson-distribution. A deviating super-Poisson behavior has been reported e.g. in single-electron tunneling through quantum dots \cite{frickePRB2007} and in single-electron tunneling at high magnetic fields \cite{kurzmannPRL2019}. Early theory for the full counting statistics in the case of incoherent tunneling and without any coherent quantum dynamics was given by Belzig \cite{belzigPRB2005}. These examples considered the property of only one observable. The correlation of two observables, namely the single electron dynamics in a quantum dot and a probe current through an adjacent quantum point contact, was reported in 2007 \cite{sukhorukovNATPHYS2007}.   
 Depending on the problem, classical rate equations or the so-called $n$-resolved master equation have been used to calculate cumulants of the counting statistics \cite{flindtPNAS2009,cookPRA1981}, factorial cumulants \cite{kamblyPRB2011,stegmannPRB2015}, or second- and third-order spectra of the frequency-resolved counting statistics \cite{emaryPRB2007}. 
All these approaches to characterizing quantum transport dynamics assume and require a strong continuous quantum measurement where the quantum system is immediately forced to reveal its state of occupation. Consequently, coherent quantum mechanical superpositions of the two alternatives of an occupied and unoccupied quantum dot are always destroyed by the measurement. 

The other limit of a quantum measurement, a weak continuous measurement, is, e.g., realized by spin noise spectroscopy which has been demonstrated on ensembles of spins in gases, semiconductors, and even on single spins \cite{crookerNATURE2004,oestreichPRL2005,dahbashiPRL2014}.
 Here, the Faraday-rotation of a probe laser beam is measured to reveal spin fluctuations \cite{hubnerPSSB2014}. Owing to the weak measurement, the spins are not projected onto spin eigenstates but may coherently process in an external magnetic field. The power spectrum of the time-resolved Faraday-signal $z(t)$ reveals a peak at the precession frequency and a broad background due to Gaussian shot noise of the probe laser.
Spin noise theories of the power spectrum have been given in terms of the spin-spin correlation function \cite{braunPRB2007}, 
Langevin approaches \cite{glazovPRB2012}, or path integral methods for weak quantum measurements \cite{sinitsynRPP2016,bednorzNJP2012}.

The purpose of this article is to demonstrate that the evaluation of both strong and weak measurement regimes can be unified in a common framework also including the intermediate regime. While transport experiments in the intermediate regime have been reported, their evaluation had been limited to second-order spectra and was lacking comparison with theory \cite{kungENTROPY2010}. The framework here is based on frequency-resolved higher-order correlation functions (so-called polyspectra \cite{Brillinger1965}) of the detector output $z(t)$. Only recently, compact quantum mechanical expressions were found for quantum polyspectra up to fourth order from continuous measurement theory \cite{hagelePRB2018,hagelePRB2020E}. 
We will treat a real-world example from nano-electronics where we use polyspectra to characterize the stochastic measurement traces and compare them with quantum polyspectra calculated from the stochastic master equation  \cite{jacobsCP2006}. As a result, we obtain important parameters of the system like tunneling times and spin relaxation rates. 
An efficient method for evaluating quantum mechanical expressions of quantum polyspectra is described in Appendix \ref{app:QuantumPolyspectra}. The formulas required for the calculation of higher-order spectra from experimental time traces can be found in Appendix \ref{app:polyspectra} including cumulant estimators and correct usage of spectral window functions.

 The stochastic master equation (SME) is an approach to continuous quantum measurements that provides a stochastic differential equation for both the system's density matrix and the detector output $z(t)$ \cite{korotkovPRB1999,korotkovPRB2001,barchielliBOOK2009,jacobsCP2006}. The coherent evolution, environment-induced damping in Markov approximation, the detector output, a stochastic measurement-induced backaction on the system, and a measurement-induced damping (Zeno-effect) are modeled. Thus, the SME is able to unify the full regime from weak to strong measurements.	
 We, therefore, consider the SME a very general most direct link between the measurable quantity $z(t)$ and the properties of the quantum system which enter the master equation. As $z(t)$ can in principle be fully characterized in terms of multi-time moments $\langle z(t_n) .. z(t_1) \rangle$, an uncompromising approach to its evaluation requires quantum mechanical expressions for such moments.
While multi-time correlators of detector output had been derived in many different contexts of varying generality [see e.g. \cite{zoller1997, bednorzNJP2012} and discussion after Eq. (\ref{eq:MomentsK})], a direct derivation of multi-time moments from the very general SME had been found by three different groups independently only in 2018 \cite{atalayaPRA2018,tilloyPRA2018,hagelePRB2018}.
This paved the way for finding compact expressions for \mbox{second-,} third-, and fourth-order cumulants as well as their corresponding quantum
 polyspectra and developing recipes for an efficient numerical evaluation \cite{hagelePRB2018,hagelePRB2020E}.  
 We use the term "quantum polyspectra" as recently introduced by Wang for polyspectra of the detector output of continuous quantum measurements \cite{wangPRR2020}.  
 Roughly speaking, the polyspectra of $z(t)$ can be interpreted as $n$th order correlators of its
  Fourier-coefficients $a_\omega$ (see App. \ref{app:polyspectra} for a strict definition). The usual powerspectrum $S^{(2)}(\omega)$ is then
   given by the expectation value $\langle a^*_\omega a_\omega \rangle$ and thus by the average intensity of $z(t)$ at frequency $\omega$. The third-order spectrum
 $S^{(3)}(\omega_1, \omega_2)$ (often called the bispectrum) is strongly related to
  $\langle a_{\omega_1} a_{\omega_2} a^*_{\omega_1 + \omega_2} \rangle$ and is sensitive to time-inversion
   (while $S^{(2)}$ is not) \cite{efimovQJRMS2001}. The fourth-order spectrum (trispectrum) usually depends on three frequencies. Below, 
   we will only consider a two-dimensional cut which is related to 
   $\langle a^*_{\omega_1} a_{\omega_1} a^*_{\omega_2} a_{\omega_2} \rangle -  \langle a^*_{\omega_1} a_{\omega_1} \rangle \langle a^*_{\omega_2} a_{\omega_2} \rangle $. The spectrum $S^{(4)}(\omega_1,\omega_2)$ may therefore be interpreted as a frequency-dependent intensity-intensity correlation. 
Emary {\it et al.} gave an early example of a bispectrum related to transport theory of quantum dots  \cite{emaryPRB2007}. Their bi\-spec\-trum for the current through a quantum dot follows from the $n$-resolved master equation which requires the strong measurement limit. Moreover, most experiments do not access the current from the quantum dot but its occupation. A recent example of a measured bispectrum of a {\it current} was therefore reconstructed from an {\it occupation} measurement \cite{ubbelohdeNATCOMM2012}. Here, we directly evaluate occupation measurements via their polyspectra.

\section{Telegraph Signal from a Single Quantum Dot}
\label{2}
\begin{figure*}[t]
  	\centering
  	\includegraphics[width=16cm]{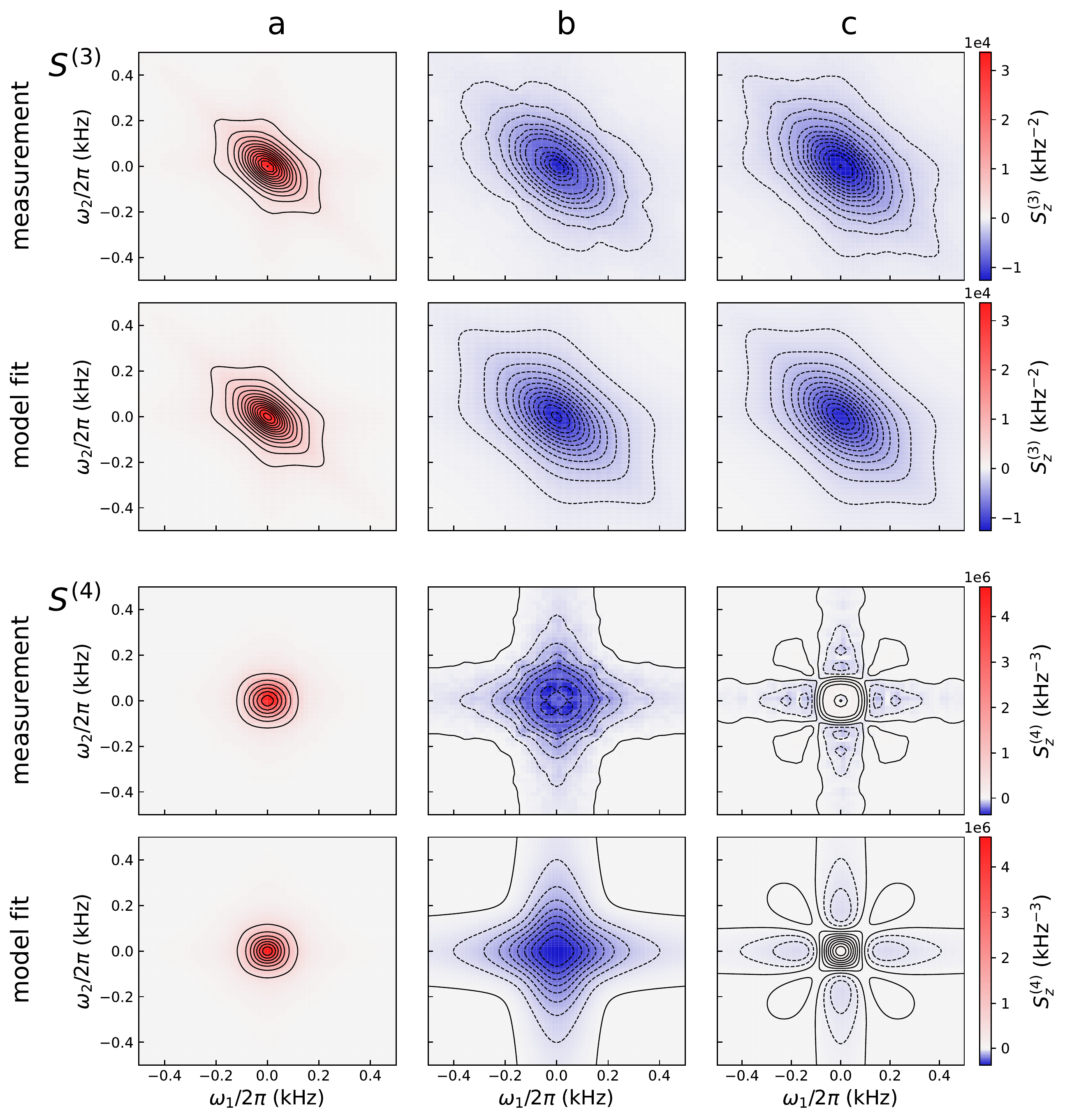}
  	\caption{(upper two rows) Experimental bispectra $S^{(3)}_z(\omega_1, \omega_2)$ at $B = $ 10 T and model fits for the tunnel regimes a) (360 mV), b) (380 mV), and c) (382 mV) (compare Fig. \ref{sample_s2}). (lower two rows) Experimental trispectra $S^{(4)}_z(\omega_1, \omega_2)$ and model fits. The analytical expressions of the polyspectra reveal that the bispectra are sensitive to the sign of $\gamma_{\text{out}}-\gamma_{\text{in}}$, whereas the trispectrum depends mostly on $(\gamma_{\text{out}}-\gamma_{\text{in}})^2$.}
  	\label{sample_poly}   
  \end{figure*}

The time traces we are going to model and evaluate were recorded in an experiment by Kurzmann {\it et al.} \cite{kurzmannPRL2019}. 
A single InAs quantum dot within an electrically biased quantum dot layer in a GaAs-based p-i-n diode structure is optically read out via resonance
fluorescence (see Fig. 1). The fluorescence time traces $z(t)$ exhibit telegraph noise due to the electron occupation dynamics of the quantum dot.
The inset of Fig. \ref{sample_s2} shows traces for an external magnetic field of 10~T and different gate voltages labeled with a) (360 mV), b) (380 mV), and c) (382 mV). The gate voltages shift the chemical potential of the electron reservoirs with respect to the single-electron level resulting in voltage-dependent tunnel rates. Corresponding power 
spectra $S_z^{(2)}(\omega)$ were calculated from time traces of 6 minutes duration each 
(evaluation scheme see App. \ref{app:polyspectra}). 
While the traces clearly show quantitative differences
in, e.g., typical up- and down-times, the power spectra are all Lorentzian-shaped and differ only weakly in widths and overall height (see Fig. \ref{sample_s2}). 
This changes dramatically for the bispectrum $S^{(3)}(\omega_1,\omega_2)$ and the fourth-order correlation 
spectrum $S^{(4)}(\omega_1,\omega_2)$ (definitions see App. \ref{app:polyspectra}). Now, clear differences become visible (see Fig. \ref{sample_poly}):
The bispectrum $S^{\rm (3)}$ is completely positive for case a) while it is negative for cases b) and c) showing a similar overall structure. Clear differences between cases b) and c) are, however, found in the trispectrum which exhibits a negative peak for case b) but an almost flat negative structure for case c).
  \label{master}

In the following, we formulate the SME for the single dot experiment described by Kurzmann. The QD dynamics is modeled by an electron tunneling rate $\gamma_{\rm in}$ onto the dot and 
a rate $\gamma_{\rm out}$ from the dot. Spin-dependent effects can be neglected at high magnetic fields (see below). The quantum states related to the occupied and empty dot follow from each other via a
 fermionic creation operator $a^\dagger$ and an annihilation operator $a$, respectively. The number operator $n = a^\dagger a$ assumes the eigenvalues 1 and 0, respectively. 
The SME propagates the density matrix $\rho(t)$ of the quantum dot 
 while it is constantly monitored for its occupation by a continuous measurement given by the operator $A = n$. 
 This simple model will be extended to a spin-dependent version in Sec. \ref{three_level_spectra} covering the case of external fields below 10~T.  
The stochastic master equation (Ito-calculus)
	\begin{eqnarray}\label{eq:sme}
	\mathrm{d}\rho &=& \frac{i}{\hbar}[\rho, H]\mathrm{d}t + \frac{\gamma_{\rm in}}{2} \mathcal{D}[a^\dagger](\rho) \mathrm{d}t + 
	\frac{\gamma_{\rm out}}{2} \mathcal{D}[a](\rho) \mathrm{d}t \nonumber \\
	&+& \beta^2 \mathcal{D}[A](\rho) \mathrm{d}t + \beta \mathcal{S}[A](\rho)\mathrm{d}W \\
	&=& \mathcal{L}[\beta] (\rho) \textrm{d}t + \beta \mathcal{S}[A](\rho)\textrm{d}W
	\end{eqnarray} 
with damping terms
	\begin{equation}
	\mathcal{D}[c](\rho) =  c\rho c^\dagger - (c^\dagger c \rho + \rho c^\dagger c)/2 ,
	\end{equation}
and backaction term
	\begin{equation}
	\mathcal{S}[c](\rho) = c \rho + \rho c^\dagger - \mathrm{Tr}\left[ (c + c^\dagger)\rho \right] \rho
	\end{equation}
describes the system dynamics $\rho(t)$ and the resulting detector output	
\begin{equation}
	z(t)  =  \beta^2 {\rm Tr}[\rho(t) (A + A^\dagger)/2]+\frac{1}{2} \beta \Gamma(t) \label{SME_detector}  
\end{equation}

as it is monitored for the measurement operator $A$ with measurement strength $\beta$.
We use the notation for the measurement strength $\beta$ from Ref.   \onlinecite{hagelePRB2018}  and for damping ${\cal D}$ 
and backaction ${\cal S}$  from Ref. \onlinecite{tilloyPRA2018}.  The SME has been derived in various forms and varying generality and was rediscovered several times in literature  \cite{barchielliNC1982,barchielliBOOK2009,belavkinConf1987, diosiPLA1988,gagenPRA1993,korotkovPRB1999,korotkovPRB2001,goanPRB2001}.  
An especially intuitive way of deriving the SME was given by Gross {\it et al.} \cite{Gross2018} and similarly by Atal {\it et al.} \cite{Attal2006, Attal2010}. They  introduce a continuous sequence of two-level quantum systems (qubits) that each weakly interact for a short period with the system and are subsequently readout by a projection measurement. The projection measurements give some information on the system and 
at the same time result in Gaussian background noise as the measurement outcome is still highly stochastic. The interaction with the stream of probe systems also causes measurement-induced system damping.
The first line of Eq. (\ref{eq:sme}) is identical to a von-Neumann master equation in Lindblad form which correctly describes incoherent tunneling on and from the quantum dot \cite{lindbladSPRINGER1976}. The Hamiltonian $H$ is set to zero since the model here disregards spin dynamics or other coherent behavior. The second line contains the differential of a stochastic Wiener process $\mathrm{d}W$
where $\Gamma(t) = \mathrm{d}W/\mathrm{d}t$ with $\langle \Gamma(t) \Gamma(t') \rangle = \delta(t-t')$ is delta-correlated Gaussian noise.
Constant monitoring of occupation $n$ leads to overall damping towards an eigenstate of $n$ (first term of the second line) and a stochastic measurement backaction on the system 
which is correlated with the detector output $z(t)$ [Eq. (\ref{SME_detector})] via the common Wiener process. 
The SME can in principle be solved numerically to simulate time traces $z(t)$ (see Fig. \ref{weak_traces}) which can then be evaluated in terms of polyspectra. Instead, we will use general analytical expressions for the 2nd- to 4th-order multi-time cumulants of $z(t)$ for general systems to obtain  expressions 
for higher-order spectra of the quantum dot dynamics \cite{hagelePRB2018,hagelePRB2020E}. The expressions are given in terms of the system Liouvillian ${\cal L}[\beta](\rho)$ where
${\cal L}[\beta](\rho)$ covers all RHS terms of Eq. (\ref{eq:sme}) but the stochastic backaction term which is non-linear in $\rho$. We define a system propagator ${\cal G}(\tau) = e^{{\cal L} \tau}\Theta(\tau)$ with the Heaviside-stepfunction $\Theta(\tau)$, a steady state $\rho_0 =   {\cal G}(\infty) \rho(t)$,
a measurement operator $A$, and its corresponding super operator \cite{noteAdagger} ${\cal A} x = (A x + x A^\dagger)/2$. 
These
definitions allow for a compact notation of multi-time moments \cite{atalayaPRA2018,tilloyPRA2018,hagelePRB2018}
\begin{eqnarray}
  \langle z(t_n) &&\cdots z(t_1) \rangle = \nonumber\\
  \beta^{2n} &&{\rm Tr}[{\cal A}{\cal G}(t_n - t_{n-1}){\cal A} \cdots {\cal G}( t_{2}-t_1) {\cal A}\rho_0 ],
  \label{eq:MomentsK}
\end{eqnarray}
where time order $t_n > t_{n-1} > ... > t_1$ is required and the system is assumed to be in its steady state $\rho_0$. Consequently, the moments depend only on time differences but not on absolute times. 
Quantum mechanical expressions for multi-time moments in the form of Eq. (\ref{eq:MomentsK}) have been given in the literature before for several special cases. Zoller and Gardiner discuss moments of the photon counting statistics [see Ref. \onlinecite{zoller1997}, Eq. (98)].
Similarly, already in 1981 Srinivas gives an early theory for photon counting where expressions with very similar structure compared to Eq. (\ref{eq:MomentsK}) appear \cite{srinivasOA1981}.
Bednorz {\it et al.} derive a moment generating functional within a path integral theory  assuming a weak measurement limit and evaluate the functional to arrive at  Eq. (\ref{eq:MomentsK})   [see Ref. \onlinecite{bednorzNJP2012}, Eq. (17)]. Wang and Clerk find the same functional as Bednorz via a Keldysh approach and use it to calculate "Keldysh-ordered" moments, cumulants, and spectra of quantum noise up to third order [Ref. \onlinecite{wangPRR2020}, Eq. (3)].  Nazarov’s full counting statistics is based on the Keldysh approach and uses a coupled ancilla system to derive a generating functional for moments
 that yields Keldysh order operators in the Heisenberg picture. The dependence 
 of detector back-action on the measurement strength is however not regarded in that approach \cite{nazarovEJP2003,nazarovBOOK2003}.

Jordan and coworkers give a path integral framework for treating continuous quantum measurements and apply it to
the simultaneous continuous measurement of two non-commuting observables of a single qubit \cite{chantasriPRA2013,chantasriPRA2015}. They, however, derive  "self-correlators" only up to second order \cite{chantasriPRA2018}. Jordan and coworkers also give an early example of a violation of a
generalized Leggett-Garg inequality which is based on second-order moments of weak measurements \cite{jordanPRL2006,williamsPRL2008}.  
Below we will see that a polyspectrum of at least third order is required to extract tunneling rates from quantum dot measurements. The great advantage of using the SME is that it provides a solid foundation for deriving analytical expressions for higher-order multi-time moments without restrictions on the measurement strength as well as a way to simulate experiment-like time-traces $z(t)$ (see Sec. \ref{sec:weak}).

 Cumulants instead of moments are often used in statistics since cumulants of the sum of {\it independent} stochastic variables are simply the sum of the individual cumulants. Additive noise in a measurement can therefore be subtracted from cumulant-based quantities. Consequently, the cumulant-based polyspectra are the desired quantities for evaluating quantum noise time traces (App. \ref{app:polyspectra}). A modified propagator ${\cal G}'(\tau) = {\cal G}(\tau)-  {\cal G}(\infty)\Theta(\tau)$ and a modified measurement super operator      ${\cal A}' x = {\cal A} x
 - {\rm Tr}({\cal A}\rho_0)x$ allow for a compact notation of multi-time cumulants \cite{hagelePRB2018,hagelePRB2020E} despite their 
 generally intricate representation in terms of moments (App. \ref{app:Cumulants}).
 The expressions
	\begin{widetext}
	\begin{eqnarray}
	C_2(z(t_1),z(t_2)) & = & \frac{\beta^2}{4} \delta(t_2-t_1) + \beta^4\sum_{\text{prm. $t_j$}} {\rm Tr}[{\cal A}' {\cal G}'(t_2 - t_1){\cal A}'\rho_0], \\
	 C_3(z(t_1),z(t_2),z(t_3)) & = & \beta^6 \sum_{\text{prm. $t_j$}}  {\rm Tr}[{\cal A}' {\cal G}'(t_3 - t_2){\cal A}'{\cal G}'(t_2 - t_1){\cal A}'\rho_0], 
	\label{eq:C3}\\
	 C_4(z(t_1),z(t_2),z(t_3),z(t_4)) & = & \beta^8 \sum_{\text{prm. $t_j$}} {\rm Tr}[{\cal A}' {\cal G}'(t_4 - t_3){\cal A}'{\cal G}'(t_3 - t_2){\cal A}'{\cal G}'(t_2 - t_1){\cal A}'\rho_0] \nonumber \\
	 &-& \beta^8 \sum_{\text{prm. $t_j$}} {\rm Tr}[{\cal A}' {\cal G}'(t_4 - t_3){\cal G}'(t_3 - t_2){\cal A}'\rho_0]{\rm Tr}[{\cal A}' {\cal G}'(t_3 - t_2){\cal G}'(t_2 - t_1){\cal A}'\rho_0] \nonumber \\
	 &-& \beta^8 \sum_{\text{prm. $t_j$}} {\rm Tr}[{\cal A}' {\cal G}'(t_4 - t_3){\cal G}'(t_3 - t_2){\cal G}'(t_2 - t_1){\cal A}'\rho_0]{\rm Tr}[{\cal A}' {\cal G}'(t_3 - t_2){\cal A}'\rho_0] \label{eq:C4}
	\end{eqnarray}
\end{widetext}
are also valid for equal times and hold without any restrictions on the time order \cite{hagelePRB2018}. The term under the sum yields a contribution if and only if the correct time order for $t_1$ to $t_4$ 
is fulfilled by one of the permutations.  The delta-function in $C_2$ appears due to the Gaussian noise  contribution $\Gamma(t)$ to $z(t)$ which is delta-correlated [Eq. (\ref{SME_detector})]. Unlike moments, cumulants beyond second order are not sensitive to Gaussian noise explaining the absence of delta-functions in $C_3$ and
$C_4$.\mbox{ Compact expressions }for cumulants beyond the fourth order are still elusive.

Next, we calculate the multi-time cumulants and quantum polyspectra for the quantum dot model, Eq. (\ref{eq:sme}).	
The Liouvillian ${\cal L}[\beta]$ of the quantum dot system can be represented as a $4\times4$ matrix with a relatively simple structure acting on the density matrix which
itself can be 
represented by a vector with four entries (compare Section XV of Ref. \onlinecite{hagelePRB2018}). The quantities $e^{{\cal L}\tau}$ and $\rho_0$
can be expressed analytically with the help of computer algebra.
Assuming time order $t_4 > t_3 > t_2 > t_1$, we find the cumulants
\begin{widetext}
	\begin{eqnarray}
	C_2(z(t_1),z(t_2)) &=& \beta^4 \frac{\gamma _{\text{in}} \gamma _{\text{out}} e^{-\left(\gamma _{\text{in}}+\gamma _{\text{out}}\right)\tau_1 }}{\left(\gamma _{\text{in}}+\gamma _{\text{out}}\right){}^2} + \frac{\beta^2}{4} \delta(t_2-t_1),
\end{eqnarray}
\begin{eqnarray}		
	 C_3(z(t_1),z(t_2),z(t_3)) &=& \beta^6 \frac{\gamma _{\text{in}} \gamma _{\text{out}} \left(\gamma _{\text{in}}-\gamma _{\text{out}}\right) e^{- \left(\gamma _{\text{in}}+\gamma _{\text{out}}\right)(\text{$\tau_1 $}+\text{$\tau_2$})}}{\left(\gamma _{\text{in}}+\gamma _{\text{out}}\right){}^3}, 
\end{eqnarray}
\begin{eqnarray}	 
	 C_4(z(t_1),z(t_2),z(t_3),z(t_4)) &=& \beta^8 \frac{\gamma _{\text{in}} \gamma _{\text{out}} \left(\left(\gamma _{\text{in}}-\gamma _{\text{out}}\right){}^2 e^{ \left(\gamma _{\text{in}}+\gamma _{\text{out}}\right)\text{$\tau_2 $}}-2 \gamma _{\text{in}} \gamma _{\text{out}}\right) e^{-\left(\gamma _{\text{in}}+\gamma _{\text{out}}\right) ((\text{$\tau_1$}+2 \text{$\tau_2$}+\text{$\tau_3 $}))}}{\left(\gamma _{\text{in}}+\gamma _{\text{out}}\right){}^4},
	 \label{eq:c4dot}
	\end{eqnarray}
where we have introduced the positive time differences $\tau_{i} = t_{i+1} - t_{i}$.
Apart from the $\beta$-prefactors and the delta-function contribution to $C_2$, the expressions $C_2$ and $C_3$ agree with those derived 
from a classical rate equation model \cite{liNJP2013}. The cumulants $C_2$ to $C_4$ show no further dependence on $\beta$ despite the fact that the Liouvillian ${\cal L}[\beta]$ contains a $\beta$-dependent damping term. A dependence of cumulants on $\beta$ can be found, e.g., in systems where a large measurement strength $\beta$ leads to suppression of coherent dynamics (Zeno-effect) \cite{misraJMP1977,korotkovPRB2001}. The absence of a $\beta$-dependence in our system is explained by the absence of coherent dynamics ($H = 0$).

The analytical expressions for the polyspectra follow after Fourier transformation of the cumulants with respect to $t_j$ [Eq. (\ref{eq:defPolyspectra})].
In Sec. XIV of Ref. \onlinecite{hagelePRB2018} it is shown how in general the time order can be dealt with when performing a multi-dimensional Fourier transform. We obtain
	\begin{eqnarray}
	S^{(2)}_z(\omega) &=& \beta^4\frac{ 2 \gamma _{\text{in}} \gamma _{\text{out}}}{\left(\gamma _{\text{in}}+\gamma _{\text{out}}\right) \left(\left(\gamma _{\text{in}}+\gamma _{\text{out}}\right){}^2+\omega ^2\right)} + \frac{\beta^2}{4},\label{SQD-s2} \\
	S^{(3)}_z(\omega_1, \omega_2) &=&\beta^6 \frac{2\gamma _{\text{in}} \gamma _{\text{out}} \left(\gamma _{\text{out}}-\gamma _{\text{in}}\right) \left(3 \left(\gamma _{\text{in}}+\gamma _{\text{out}}\right){}^2+\omega _1^2+\omega _2^2+\omega _1 \omega _2\right)}{  \left(\gamma _{\text{in}}+\gamma _{\text{out}}\right) \left(\left(\gamma _{\text{in}}+\gamma _{\text{out}}\right){}^2+\omega _1^2\right) \left(\left(\gamma _{\text{in}}+\gamma _{\text{out}}\right){}^2+\omega _2^2\right) \left(\left(\gamma _{\text{in}}+\gamma _{\text{out}}\right){}^2+\left(\omega _1+\omega _2\right){}^2\right)}.
	\label{SQD-s3}
	\end{eqnarray}
	\end{widetext}
A cut through the trispectrum $S_z^{(4)}(\omega_1,\omega_2) = S_z^{(4)}(\omega_1,-\omega_1, \omega_2)$ is given in the Appendix, Eq. (\ref{eq:s4xx}). 
Alternatively, a direct evaluation of the quantum polyspectra  in the frequency domain yields the 
same results (App. \ref{app:QuantumPolyspectra}). Similar to the cumulant expressions, the measurement strength $\beta$ leads for the present system only to a prefactor in the polyspectra and a variation of the spectrally flat  background noise in  $S_z^{(2)}$.  In contrast, we will treat in Section \ref{sec:double_dot} a double-dot system where an increasing measurement strength $\beta$ leads to a clear suppression of coherent quantum oscillations and a corresponding overall change in the polyspectra.

The experimental time traces of the quantum dot occupation dynamics are evaluated for  $\gamma_{\rm in}$ and $\gamma_{\rm out}$ by simultaneously fitting the analytical expression of the spectra $S_z^{(2)}(\omega)$, $S_z^{(3)}(\omega_1,\omega_2)$,
and $S_z^{(4)}(\omega_1,\omega_2)$ to the spectra of the measured time traces. A constant background contribution to the power spectrum $S_z^{(2)}(\omega)$ is also regarded separately. Time traces of 6 min durations with a temporal resolution of 100~$\mu$s were 
taken for each gate voltage. The measured traces and $z(t)$ differ by a setup-dependent scaling factor that is regarded in the fitting procedure.  In our case, the scaling factor is negative since the occupied quantum dot state results in absent fluorescence.

\begin{figure}[t]
 	\centering
 	\includegraphics[width=7.3cm]{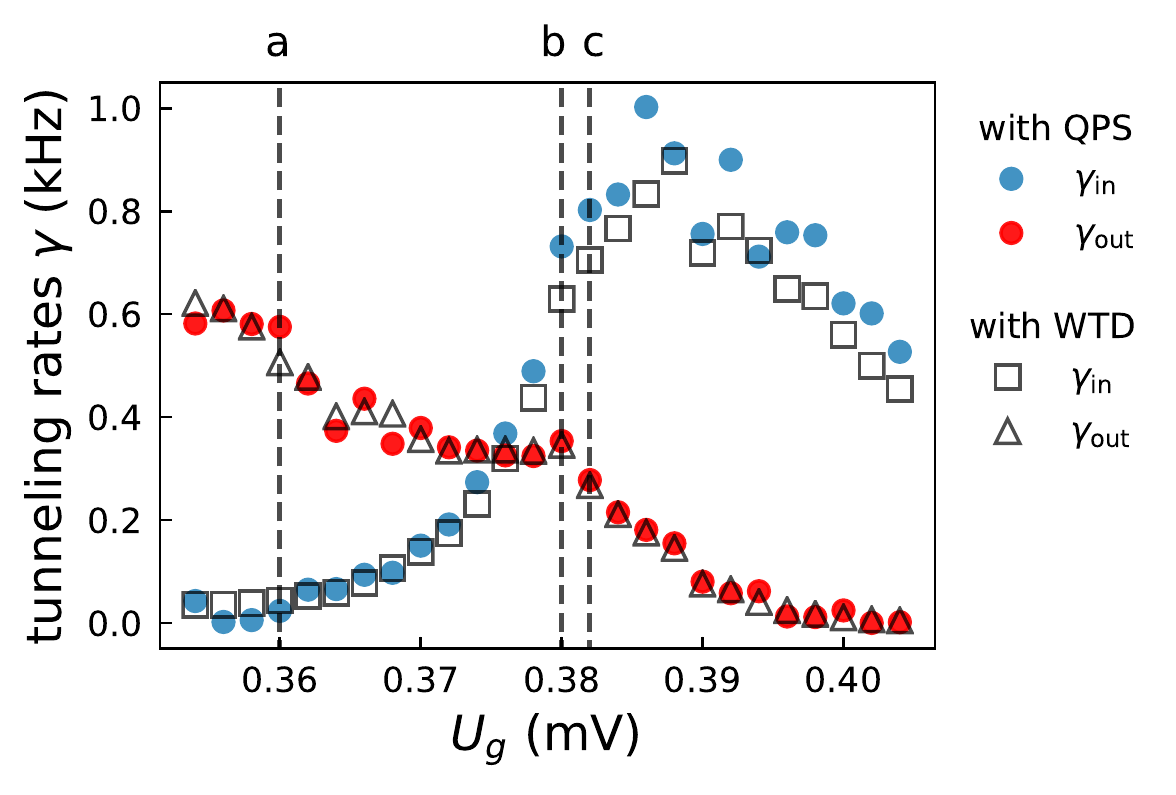}
 	\caption{Tunneling rates determined from experimental polyspectra of the full time trace (blue and red dots) in comparison with values obtained from  the waiting-time distribution  (open squares and triangles) \cite{kurzmannPRL2019}. Both methods arrive at the same results. The vertical lines mark the gate voltages belonging to the time-traces shown in Fig. \ref{sample_s2}.}
 	\label{gammas} 
 \end{figure}

 The polyspectra for cases a), b), and c) obtained from fitting are displayed for illustration along with the originally measured spectra in Fig. \ref{sample_poly}.
 The three sample traces of Fig. \ref{sample_s2} can now be attributed to three regimes of the tunneling rates: a) $\gamma_{\text{in}} \ll \gamma_{\text{out}}$, b) $\gamma_{\text{in}} \gtrapprox \gamma_{\text{out}}$, and c) $\gamma_{\text{in}} > \gamma_{\text{out}}$. 
 Figure \ref{gammas} compares as an important result the tunneling rates for different gate voltages and a magnetic field of \mbox{10 T} obtained from polyspectra with those obtained from a previous analysis of the waiting-time distribution (WTD), i.e, $p(0,t)$ of the full counting statistics \cite{kurzmannPRL2019,KurzmannSupp2019}. 
\begin{figure}[t]
	\centering
	\includegraphics[width=4.5cm]{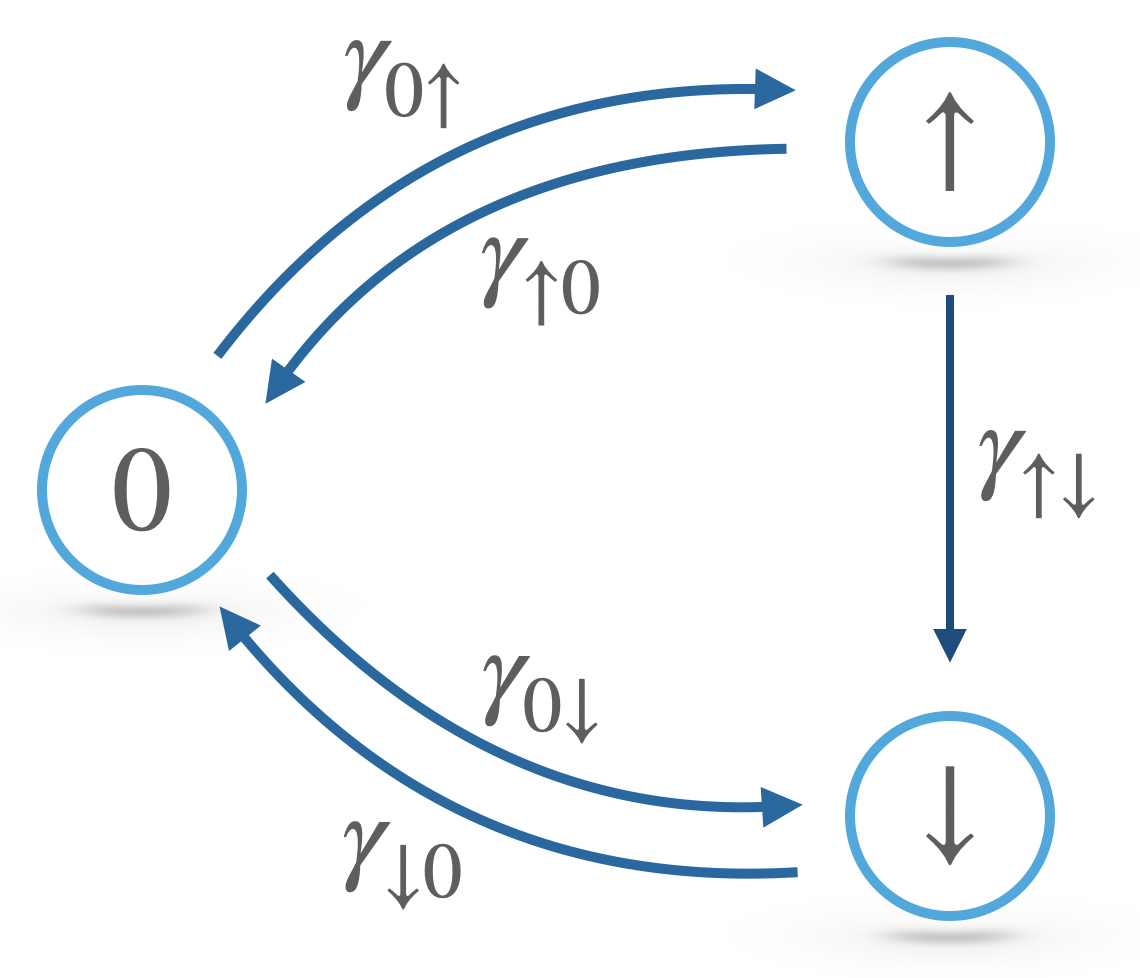}
	\caption{Three-state model of the spin-depended quantum dot dynamics.}
	\label{raten} 
\end{figure}
 The excellent agreement demonstrates that polyspectra are a powerful tool for evaluating transport measurements. 
The general weak variation in the power spectra is easily 
explained by Eq. (\ref{SQD-s2}) which does depend only on the sum $\gamma_{\text{in}} + \gamma_{\text{out}}$ of the tunnel rates. The prefactor
 $\gamma_{\text{in}} \gamma_{\text{out}}$ can not be exploited to separate  $\gamma_{\text{in}}$ and $\gamma_{\text{out}}$ since the measurement strength $\beta$ acts as an overall scaling parameter. 
In contrast, the prefactor of the bispectrum, Eq. (\ref{SQD-s3}), together with the prefactor of the 
 power spectrum contain information on  $\gamma_{\text{in}} -\gamma_{\text{out}}$ and are in principle sufficient to extract both parameters.
  We found more reliable results by simultaneously fitting also the fourth-order spectrum which is sensitive
   to $\gamma_{\text{in}} + \gamma_{\text{out}}$ and $(\gamma_{\text{in}} -\gamma_{\text{out}})^2$ [see Eq. (\ref{eq:c4dot})]. 
   We note that the second- and fourth-order spectrum do not change under exchange
    of $\gamma_{\text{in}}$ and $\gamma_{\text{out}}$ making the evaluation of the bispectrum mandatory.

\section{Spin-dependent quantum dot dynamics}
\label{three_level_spectra}
Next, we apply our method to a quantum dot in magnetic fields below 10~T. Following Kurzmann {\it et al.} \cite{kurzmannPRL2019}, the Zeeman spin splitting $\Delta$ leads to spin-dependent tunneling rates  
\begin{eqnarray}
	\gamma_{0\uparrow} &=& d \Gamma f\left( \epsilon + \Delta/2 \right), \nonumber \\
	\gamma_{0\downarrow} &=& d  \Gamma f\left( \epsilon - \Delta/2 \right), \nonumber \\
	\gamma_{\uparrow 0} &=&  \Gamma\left[  1-f\left( \epsilon + \Delta/2 \right) \right], \nonumber \\
	\gamma_{\downarrow 0} &=&  \Gamma\left[  1-f\left( \epsilon - \Delta/2 \right) \right]. \label{eq:KurzmannRates}
	\end{eqnarray}
The tunnel-coupling strength $\Gamma$  characterizes the tunnel barrier, $f(x)$ is the Fermi distribution function of the electron reservoir, the quantum dot level energy is given by $\epsilon$, and the temperature by $T = 10$~K. 
The prefactor $d=10/11$ regards a reduction of the tunneling due to the presence of the exciton whose fluorescence is detected by the measurement setup \cite{kurzmannPRL2019}. 
After introduction of a spin relaxation rate $\gamma_{\uparrow\downarrow}$ to the down state, the system is fully described by an incoherent 
transition dynamics depicted schematically in Fig. \ref{raten}. Spin flips to the up state are neglected as the spin down state is energetically favorable in 
magnetic fields \cite{kurzmannPRL2019}.
\begin{figure}[t]
	\centering
	\includegraphics[width=7cm]{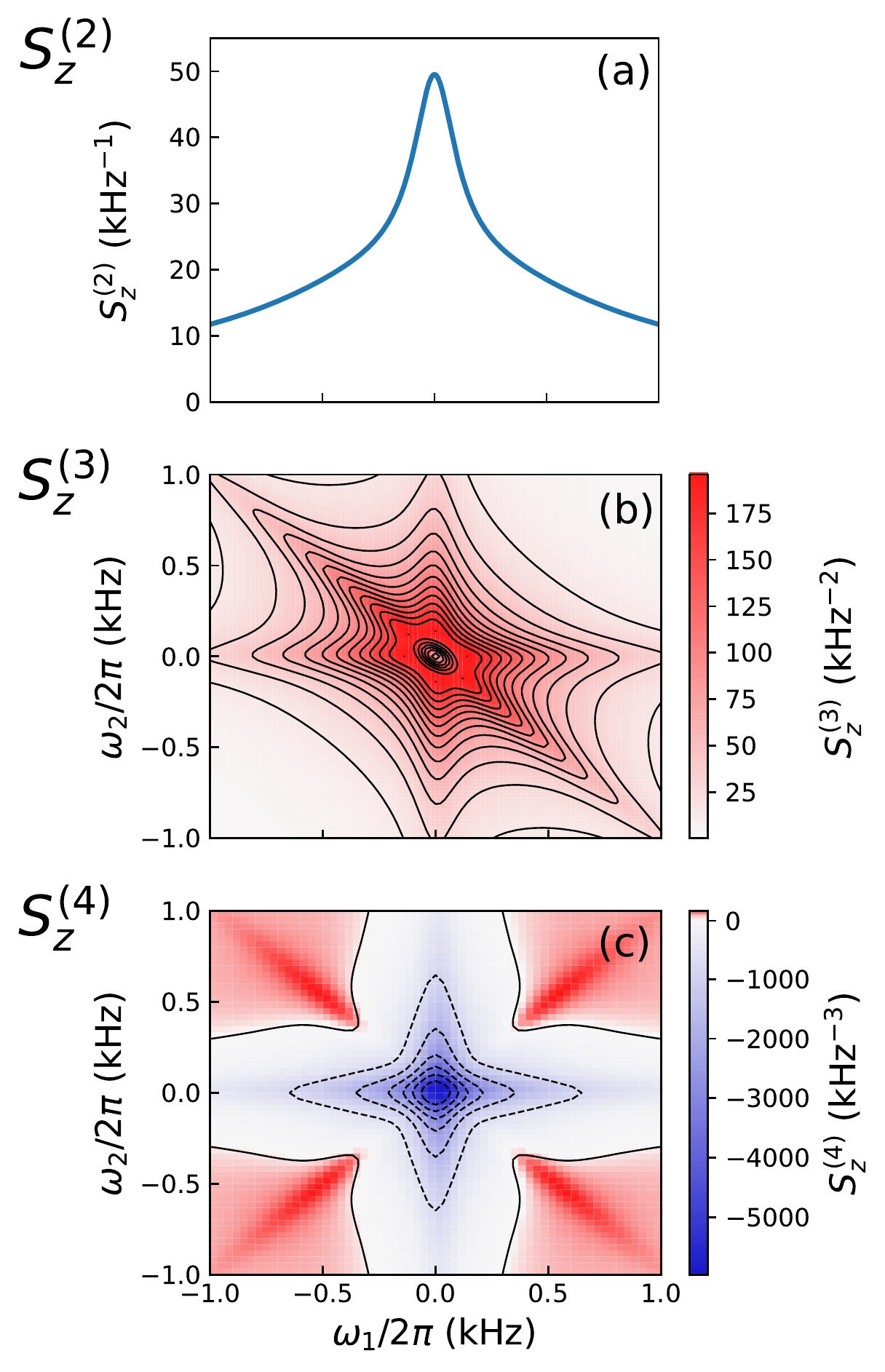}
	\caption{Calculated power-, bi- and trispectrum of the spin-dependent quantum dot dynamics for	 $\gamma_{0 \downarrow}=\gamma_{\downarrow 0}  = 2.5$~kHz,$\gamma_{0 \uparrow}=\gamma_{\uparrow 0} =0.5$~kHz, and $\gamma_{\uparrow\downarrow} = 0$ (from top to bottom). The bispectrum shows deviations from a simple Lorentzian shape and
		a dip at $\omega_1=\omega_2=0$. The trispectrum strongly deviates from those of the quantum dot at \mbox{10 T}
		(compare Fig. \ref{sample_s2} and \ref{sample_poly}).}
	\label{3level}
\end{figure}
\begin{figure}[t]
	\begin{center}
		\includegraphics[width=9cm]{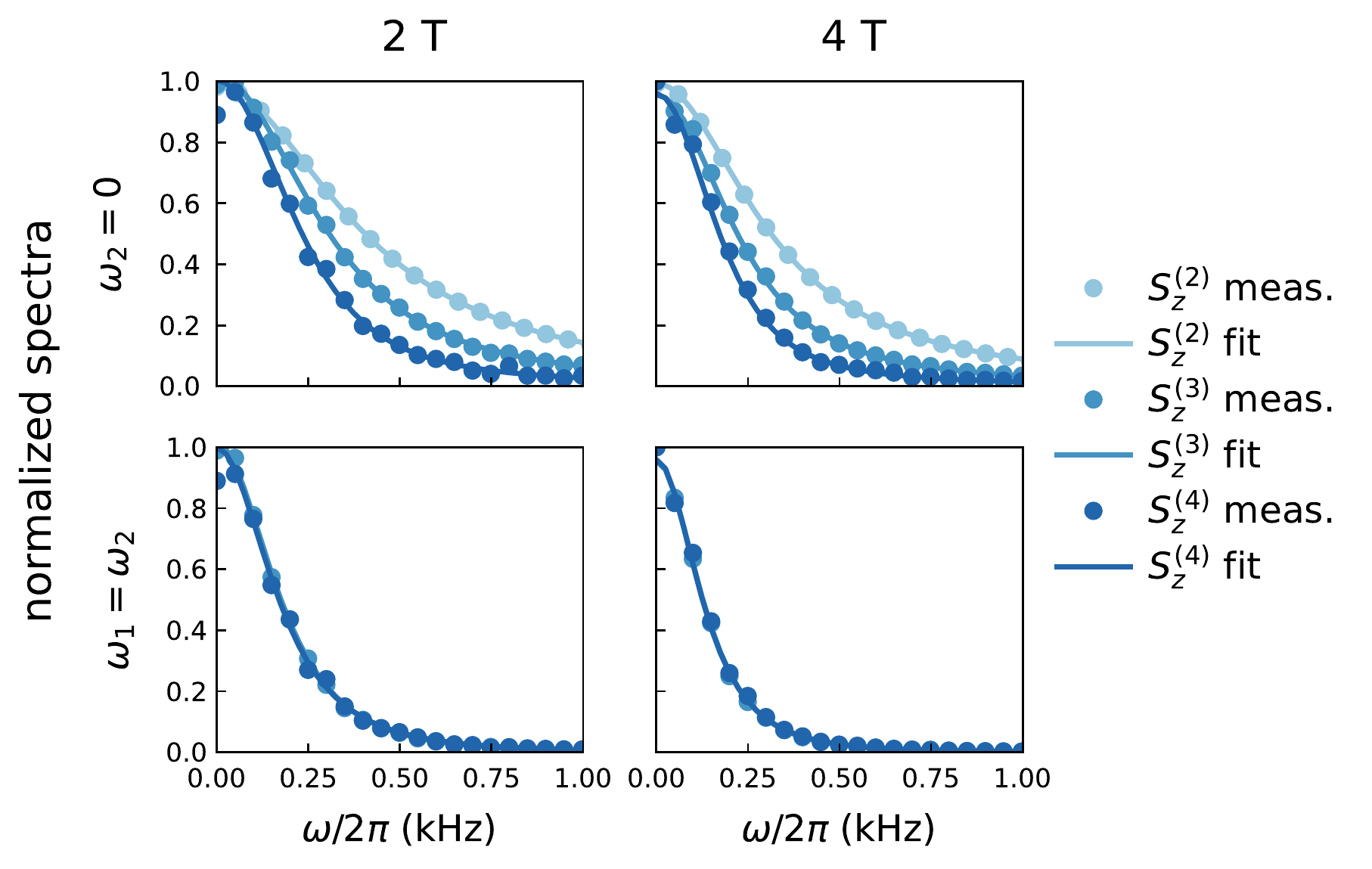}\\[3mm]
		\caption[]{Comparison between the measured and fitted polyspectra at $B =$ 2 T and 4 T. Cuts with $\omega_2 = 0$ (upper row) and cuts with $\omega_1 = \omega_2$ (lower row) are shown 
			for the bi- and trispectrum. The maximum values of all spectra have been normalized to 1. We can see excellent agreement between the measurement and the three-state model. The diagonals of the bi- and trispectrum coincidentally overlap for the quantum dot system.}
		\label{relaxation_fit}
	\end{center}
\end{figure}
The model in Kurzmann {\it et al.} can be formulated as a stochastic master equation
\begin{eqnarray}\label{eq:3_level}
\mathrm{d}\rho &=& \frac{\gamma_{0\uparrow}}{2} \mathcal{D}[a_\uparrow^\dagger a_0](\rho) \mathrm{d}t + \frac{\gamma_{0\downarrow}}{2} \mathcal{D}[a_\downarrow^\dagger a_0](\rho) \mathrm{d}t
\nonumber \\
&+&\frac{\gamma_{\uparrow 0}}{2} \mathcal{D}[a_0^\dagger a_\uparrow](\rho) \mathrm{d}t  +\frac{\gamma_{\downarrow 0}}{2} \mathcal{D}[a_0^\dagger a_\downarrow](\rho) \mathrm{d}t \nonumber \\
&+&\frac{\gamma_{\uparrow\downarrow}}{2} \mathcal{D}[a_\downarrow^\dagger a_\uparrow](\rho) \mathrm{d}t + \frac{\beta^2}{2} \mathcal{D}[n_\uparrow + n_\downarrow](\rho) \mathrm{d}t  \nonumber \\ 
&+& \beta \mathcal{S}[n_\uparrow + n_\downarrow](\rho)\mathrm{d}W,
\end{eqnarray} 
with detector output 
\begin{equation}
	z(t)  =  \beta^2 {\rm Tr}[\rho(t) (n_\uparrow + n_\downarrow)] + \frac{1}{2} \beta \Gamma(t).  
\end{equation}
The measurement operator appears as a sum  $n_\uparrow + n_\downarrow$ since the detection scheme does not distinguish between up and down spins, but is only sensitive to the mere presence of an electron in the quantum dot.

As an example, the power spectrum, bispectrum, and trispectrum were calculated from Eqs. (\ref{eq:S2}) to (\ref{eq:S4}) for 
tunneling rates $\gamma_{0\uparrow}=\gamma_{\uparrow 0}=0.5$~kHz,
$\gamma_{0\downarrow}=\gamma_{\downarrow 0} = 2.5$~kHz, and absent spin relaxation
 $\gamma_{\uparrow\downarrow} = 0$ (see Fig. \ref{3level}).  Their structure is clearly different from the spectra of the simple quantum dot model discussed above. The power spectrum appears to be a superposition of two Lorentzian peaks. The bispectrum reveals a small dip at zero frequencies and the   
trispectrum displays positive maxima on the diagonals that were absent for the simple quantum dot model.
Spin relaxation rates $\gamma_{\uparrow\downarrow}$ larger than the tunneling rate cause practically all electrons to tunnel from the QD via the spin-down level at the rate $\gamma_{\downarrow 0}$ while
 electrons enter the empty dot at an effective rate  $\gamma_{0\uparrow} + \gamma_{0 \downarrow}$. The dot dynamics, therefore, follow the simple quantum dot model and the spectra resume the appearance of spectra shown in Fig. \ref{sample_poly}. 
 The dependence of the spectra on the spin relaxation rate suggests that the spin relaxation rate can be extracted from data measured at finite magnetic fields.
This dependence will get weaker for similar tunneling rates of the two spin orientations, i.e.   $\gamma_{0\uparrow} \approx \gamma_{0 \downarrow}$ and $\gamma_{\uparrow 0} \approx \gamma_{ \downarrow 0}$.  For exact agreement, the occupation dynamics will obviously not depended on the spin orientation and is therefore not sensitive to the spin relaxation rate.

In contrast to the simple model above, analytical expressions for the spectra are not available. The fitting procedure, therefore, relies on a numerical evaluation of the quantum polyspectra via  Eqs. (\ref{eq:S2}) to (\ref{eq:S4}).  
The parameter space for the five relaxation rates is restricted by their dependence on $\Gamma$ and $\epsilon$ [compare Eqs. (\ref{eq:KurzmannRates})].    
Spin relaxation rates $\gamma_{\uparrow \downarrow}$ were determined for a gate voltage of 371~mV at a field of 2~T and for 376~mV at 4~T. We obtain an almost perfect agreement between data and model for both fits. Fig. \ref{relaxation_fit} shows for a quantitative comparison cuts of all three spectra along the $\omega_2 = 0$ axes and cuts for the bi- and trispectra along their diagonal $\omega_1 = \omega_2$. 
 \begin{figure}[t]
	\begin{center}
		\includegraphics[width=7cm]{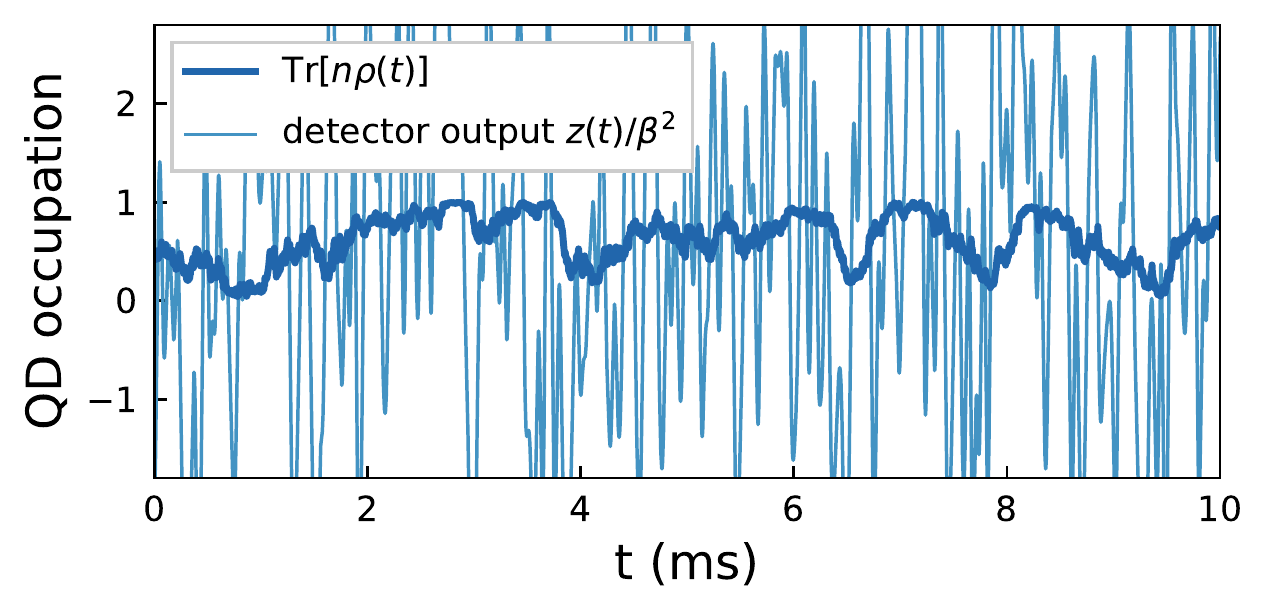}\\[3mm]
		\caption[Transient example]{Simulated detector time trace $z(t)$ of a single quantum dot from solving the stochastic master equation in the weak measurement regime (light blue line). The electron occupation ${\rm Tr}(\rho(t) n)$ (dark blue line) follows from the calculation but is not directly accessible in an experiment. The detector output has been scaled by $\beta^{-2}$ for comparison.}
		\label{weak_traces}
	\end{center}
\end{figure}
The 2~T-case yields $\gamma_{\uparrow \downarrow}^{\text{2T}} = 2.0$~kHz and tunneling rates  $ (\gamma_{0\uparrow},\gamma_{0 \downarrow},\gamma_{\uparrow 0},\gamma_{ \downarrow 0}) = (1.18,1.24,0.29,0.22)$~kHz.
For 4~T we obtain $\gamma_{\uparrow \downarrow}^{\text{4T}} = 11$~kHz and tunneling rates  $ (\gamma_{0\uparrow},\gamma_{0 \downarrow},\gamma_{\uparrow 0},\gamma_{ \downarrow 0}) = (0.89, 0.97, 0.21, 0.13)$~kHz. The discrepancy to the values given in Kurzmann {\it et al.} for 2 T $\gamma_{\uparrow \downarrow}^{\text{2T}} \approx 0.0$~kHz and 4 T $\gamma_{\uparrow \downarrow}^{\text{4T}} = 3.0$~kHz may be explained by a weak dependence of the tunneling rates on the spin orientation and a large spin relaxation (see previous paragraph). 
In both cases, the spin-relaxation $\gamma_{\uparrow \downarrow}$ rate has only little influence on the 
tunneling dynamics giving rise to large errors in the model fit. 
\begin{figure}[t]
	\begin{center}
		\includegraphics[width=8.7cm]{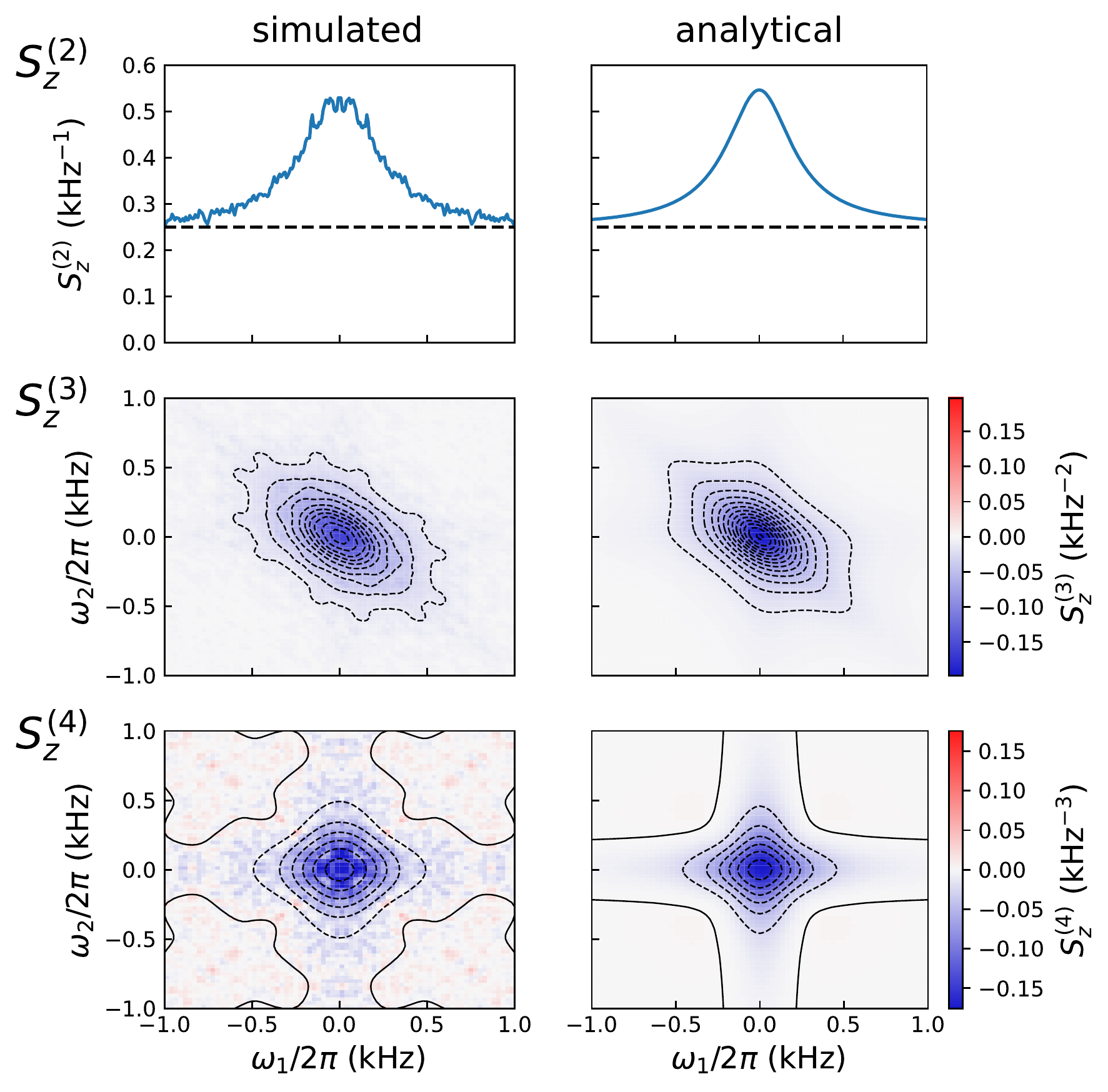}\\[3mm]
		\caption[Transient example]{Comparison between the simulated (left) and analytical (right) polyspectra of the quantum dot system in the weak measurement regime. The spectra show good agreement and  tunneling rates could successfully be recovered from the numerical spectra by a fitting procedure. The dashed lines in the power spectra indicate the level of white Gaussian background noise.}
		\label{weak_poly}
	\end{center}
\end{figure}

\section{Evaluating transport in the weak measurement regime}
\label{sec:weak}
The experimental time-traces from above exhibit telegraph noise and a small amount of additional noise. This allows in principle for the evaluation of data in terms of waiting-time distributions of the FCS. Continuous measurement theory, however, states that telegraph noise does disappear for weaker measurements.  Such a disappearance was recently reported for the case of gate-tunable quantum point contacts  \cite{kungPRB2009}. It was shown that a cross-correlation spectrum of two adjacent QPCs showed similar signatures as the spectra from
 the stronger measurement regime \cite{kungENTROPY2010}. However, an analysis in terms of the full counting statistics as, e.g., previously 
required for separating in- from out-tunneling rates is no longer possible for vanishing telegraph behavior. 
Here we show that an evaluation of general time traces in terms of quantum polyspectra is possible even in the weak measurement regime without any restrictions.

We simulate a weak measurement ($\beta^2 = 1$ kHz) on a quantum dot, Eq. (\ref{eq:sme}), with parameters
 $\gamma_{\text{in}}$ = \mbox{1 kHz} and $\gamma_{\text{out}} = 0.5$ kHz. The integration of the SME with the \textit{QuTiP} software package \cite{JOHANSSON20131234}
  yields the QD occupation ${\rm Tr}[n \rho(t)]$ (see Fig. \ref{weak_traces}, dark blue curve). It assumes values
   in the full regime between 0 and 1. Hence, the system is not being projected into one of its eigenstates. 
   The actual detector time trace exhibits large background noise (see Fig. \ref{weak_traces}, light blue curve) 
   that dramatically exceeds the interval from 0 to 1, an effect put forward by Aharonov {\it et al.} 
   in their pioneering work on the notion of weak measurements \cite{aharonovPRL1988}. 
   Quantum jumps can no longer quantitatively be evaluated from the time trace and methods related to the FCS cannot be applied.
Fig. \ref{weak_poly} (left column) shows polyspectra calculated from the simulated measurement trace with a duration of 30 minutes. The polyspectra follow from a scheme based on multivariate cumulant estimators of Fourier coefficients of the time trace (App. \ref{app:polyspectra}).   We see excellent agreement between numerical spectra and the ones evaluated from the exact expressions for quantum polyspectra, Eqs. (\ref{eq:S2}), (\ref{eq:S3}), and (\ref{eq:S4}). For comparison of spectra, no addition normalization was needed as prefactors arising, e.g., from the spectral window were
correctly accounted for by our formulas. The simulated polyspectra exhibit increasing noise for increasing order which is a known feature for estimates of cumulant-based quantities \cite{mccullaghBOOK2018}. The negative bispectrum immediately reveals $\gamma_{\text{in}} > \gamma_{\text{out}}$. A simultaneous fit of all spectra yields the predefined tunneling rate within an error of \mbox{10 \%}. Traces with even stronger background noise can be evaluated if spectra are averaged for sufficiently long measurement times.
\begin{figure*}[t]
	\centering
	\includegraphics[width=17cm]{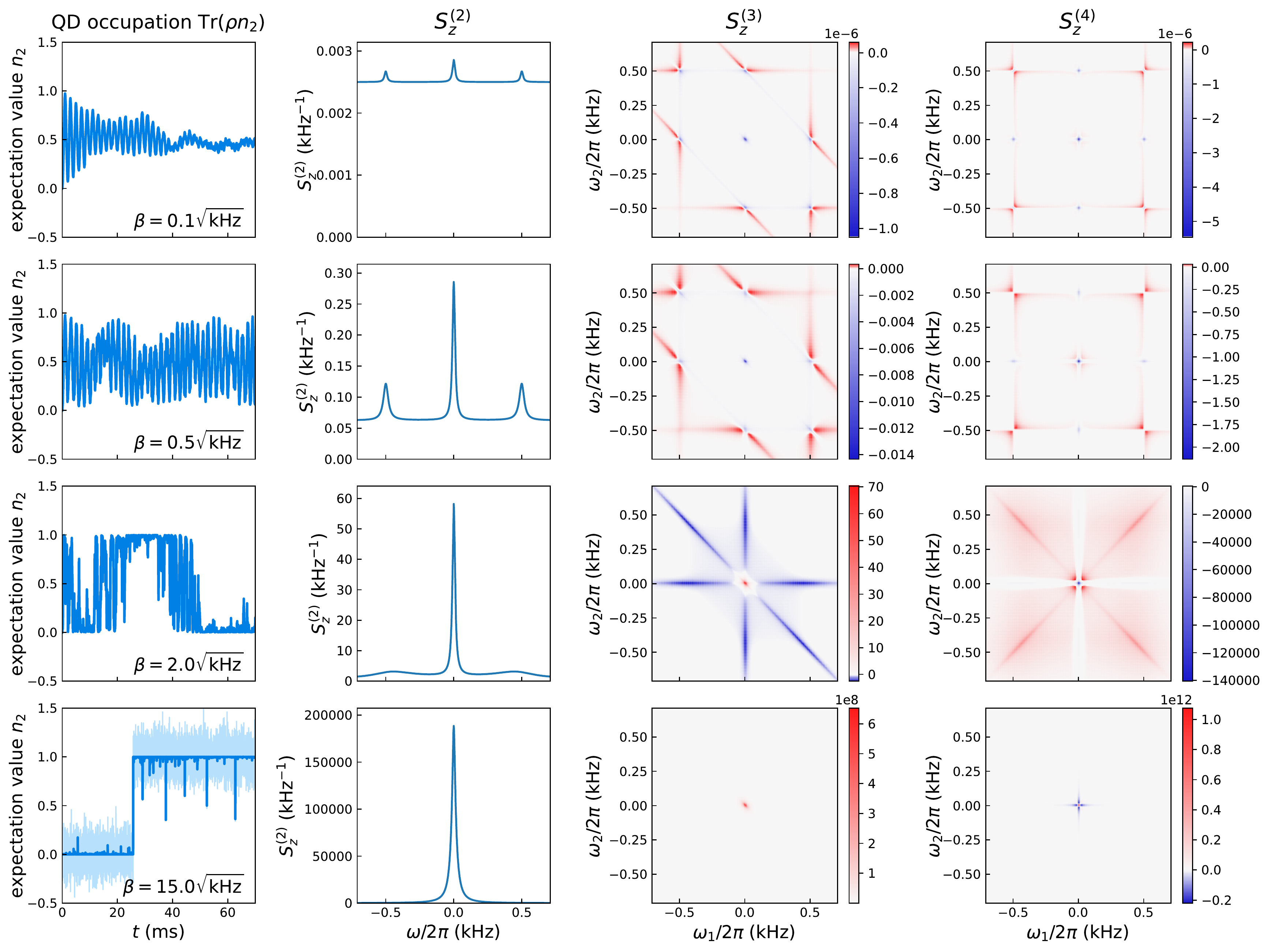}
	\caption{Occupation dynamics and quantum polyspectra of a double-dot system: the occupation dynamics of the second dot and corresponding polyspectra change for increasing measurement strengths $\beta$ ($\beta$ increases for lower rows). The coherent oscillations are suppressed for stronger measurements (quantum Zeno effect). For $\beta = \SI{2}{\sqrt{kHz}}$ the dynamics has changed to telegraph-like switching. The peak due to coherent oscillations at $\omega/2\pi = \SI{0.5}{kHz}$ in the power spectrum $S_z^{(2)}$ broadens and shifts to zero frequency (lower row). The signatures in  $S_z^{(3)}$ and $S_z^{(4)}$ are also broadened and move towards the origin. The detector output $z(t)$ is dominated by strong white Gaussian noise and therefore not displayed in the first column except for the largest measurement strength in the last row (light blue line). The units of $S^{(3)}_z$ and $S^{(4)}_z$ are kHz$^{-2}$ and kHz$^{-3}$, respectively.}  
	\label{double_dot}   
\end{figure*} 

Next, we shortly discuss the nature of the SME detector model in comparison with another model that was recently investigated by Li {\it et al.} in the context of a randomly flipping spin \cite{liNJP2013}. 
The time trace in Figure 8 corresponds to the noisy current that traverses a quantum point contact and thereby probes the quantum dot occupation. 
In the above case of weak coupling, the probe current modulation due to the dot occupation is much weaker than the omnipresent shot-noise. Nevertheless, cumulants and spectra for order three or higher exhibit only signatures of the quantum system as they are not sensitive to Gaussian shot-noise. 
Li {\it et al.} give a different example of a continuous measurement on a two-level system \cite{liNJP2013}. Their detector model yields a -1 or 1 for the two levels and 0 if no information could be gathered. For a slow telegraph dynamics of the system, the jump dynamics can be resolved in the detector stream and spectra similar to ours [Eqs. (\ref{SQD-s2}), (\ref{SQD-s3}), and (\ref{eq:s4xx})] were recovered. For system dynamics faster than the average appearance rate of 1 or -1 (a weak measurement in the terminology of Li) the authors find by a numerical simulation of the detector output that their fourth-order cumulant “quickly deteriorates” and no longer bears resemblance with the desired spectra. A full theory of their detector model that would be capable of predicting spectra in the spirit of our C1 to C3 for different measurement strength is, however, still elusive. A separation of detector-related and system-related contributions to polyspectra remains therefore as a challenge in the case of Li's detector model.

\section{Zeno-transition in a double-dot system}
\label{sec:double_dot}
The quantum system treated in the preceding section was completely dominated by incoherent tunneling dynamics and spin relaxation. Consequently, coherent quantum dynamics was neither expected to leave a signature in the experiment nor needed to be described by a Hamiltonian (i.e., so far $H = 0$).  

The SME is, however, capable of describing coherent quantum dynamics and the effect of a continuous measurement on the dynamics. Here, we investigate the transition from the weak to the strong measurement regime in a double-dot system where electrons can coherently oscillate between the dots due to tunnel coupling.  
The dynamics follows from the hopping term that quantifies the coupling of the dots
\begin{eqnarray}
H = \hbar g \left( a_1^\dagger a_2 + a_2^\dagger a_1 \right),
\end{eqnarray}
where $g$ is the coupling constant and $a_1$ ($a_2$) is the annihilation operator acting on the first (second) QD. We neglect the spin degree of the electrons. The stream of electrons between the leads through the system is modeled via incoherent tunneling as presented in Section \ref{2}. The electrons can enter the system only via the first dot and leave it only via the second dot. The Hamiltonian allows for the simultaneous occupation of both dots with one electron each. A continuous measurement of the occupation of the second quantum dot is modeled by the operator $n_2 = a_2^\dagger a_2$, which corresponds to a measurement via a QPC in the vicinity of the second dot. The SME becomes
\begin{eqnarray}\label{eq:sme_double_dot}
	\mathrm{d}\rho &=& \frac{i}{\hbar}[\rho, H]\mathrm{d}t + \frac{\gamma_{\rm in}}{2} \mathcal{D}[a_1^\dagger](\rho) \mathrm{d}t + 
	\frac{\gamma_{\rm out}}{2} \mathcal{D}[a_2](\rho) \mathrm{d}t \nonumber \\
	&+& \beta^2 \mathcal{D}[n_2](\rho) \mathrm{d}t + \beta \mathcal{S}[n_2](\rho)\mathrm{d}W 
\end{eqnarray} 
with the detector output	
\begin{eqnarray}\label{z_double_dot}
	z(t)  =  \beta^2 {\rm Tr}[\rho(t)n_2]+\frac{1}{2} \beta \Gamma(t). \label{SME_detector_double_dot}  
\end{eqnarray}
The parameters $g = \SI{\pi/2}{kHz}$, $\gamma_\text{in} = \SI{0.071}{kHz}$ and $\gamma_\text{out} = \SI{0.069}{kHz}$ are used for the calculation of system dynamics and quantum polyspectra.
Fig. \ref{double_dot} shows the transition of the system from the weak measurement regime to the quantum Zeno regime.  Each row depicts the occupation dynamics $ {\rm Tr}[\rho(t)n_2]$ and the quantum polyspectra $S^{(2)}$,  $S^{(3)}$, and $S^{(4)}$ for increasing 
measurement strengths $\beta$ [calculated from Eq. (\ref{eq:S2}), (\ref{eq:S3}), (\ref{eq:S4})].
The transients of the occupation dynamics start at $t= 0$ with the first dot fully occupied. In the weak measurement case, the system relaxes towards the steady state where  $n_1 \approx n_2 \approx 0.5$. The weak measurement is however causing non-vanishing oscillations in $ {\rm Tr}[\rho(t)n_2]$ as the system is forced to reveal partly whether the second dot is empty or full. A subsequent oscillation dynamics towards or from the dot, respectively, needs then to occur. 
The polyspectra thus reveal a signature at frequency $g / 2\pi = \SI{0.5}{kHz}$. The occupation number $n_2$ oscillates around a value that itself exhibits slow fluctuations in time. The reason being that the number of electrons in the system is not strictly conserved due to random tunneling events from and into the leads.  Consequently, a clear peak at zero frequency appears in all spectra.

For low measurement strength (first and second row), we find sharp spectral lines in $S^{(2)}$, $S_z^{(3)}$, and $S_z^{(4)}$ .  The lines in $S_z^{(4)}$ exhibits the same characteristic \textit{butterfly} signature as display in Figure 14 of Smirnov {\it et al.} \cite{smirnovARXIV2020}.
For increasing measurement strengths (third and fourth row) the oscillating dynamics become more and more suppressed.  Telegraph-like switching is observed for the strongest measurement strength where all coherent dynamics is suppressed (so-called quantum Zeno regime). Consequently, all peaks in the spectra move towards zero frequency. The incomplete switching events seen in transient for the largest measurement strength are known as spikes and have been investigated in detail by Tilloy \cite{tilloyPRA2015}. The white background noise in $S_z^{(2)}$  decreases as the expectation value in Eq. \eqref{z_double_dot} dominates over the Gaussian noise term for large $\beta$. The signatures in $S_z^{(3)}$ and $S_z^{(4)}$ are equally broadened as signals from a larger range of frequencies are getting correlated. 
The quantum Zeno effect had been previously studied theoretically by Korotkov in a double-dot system without leads. He found a damping of oscillation dynamics and a correspondingly broadened spectral line in his $S_z^{(2)}$ \cite{korotkovPRB2001}. As a consequence of leads, our $S_z^{(2)}$ exhibits an additional peak at zero frequency (see above). Korotkov finds an unrelated peak at zero frequency only after introducing an energy imbalance $\varepsilon$ of the two dots. Suppression of coherent dynamics was recently observed in experiments on a superconducting qubit in a GHz-cavity (circuit quantum electrodynamics) where measurement strengths and observation operators were highly controllable \cite{ficheuxNATCOMM2018}.
The dynamics of the system was obtained from averaging many noisy time-traces for an identically prepared initial state, while here steady state fluctuations are observed and analyzed without the need for state preparation. 
While we are  currently not fully able to interpret all features in the higher-order spectra, it is clear that they contain more information than usual power spectra. Our analysis of a real-world experiment in Sections II and III demonstrates that higher-order spectra can contribute substantially to the analysis of 
quantum systems.   

\section{conclusion}
We presented quantum polyspectra within a stochastic master equation approach as a viable alternative to full counting statistics approaches for evaluating time-traces of transport measurements. The framework is applicable to general systems where both coherent evolution and incoherent coupling to the environment are important \cite{jacobsCP2006}. The stochastic master equation has previously been shown to cover the whole regime from weak to strong measurements and allows for modeling of the weak measurement regime as well as investigating the transition to the Zeno regime \cite{korotkovPRB2001}. In quantum electronics, quantum polyspectra can in the future be used for evaluating weak quantum point contact measurements where background noise prevents a traditional analysis in terms of the full counting statistics, i.e. when quantum jumps can no longer be identified from time traces. Weakly coupled QPCs and polyspectra may be the key for fully characterizing  coherent dynamics in transport measurements like, e.g., spin precession or tunnel dynamics between adjacent quantum dots \cite{koppensNAT2006}. We also expect applications of quantum polyspectra in circuit quantum electrodynamics and quantum optics in general \cite{norrisPRL2016,ficheuxNATCOMM2018,minevNATURE2019,wangPRR2020}.

\begin{acknowledgments}
We acknowledge financial support by the Deutsche Forschungsgemeinschaft (DFG, German Research Foundation) Project No. 278162697— SFB 1242. D. H., M. G., and A.L. acknowledge financial support by the DFG by the individual research grants 
No. HA3003/7-1, No. GE2141/5-1, and No. LU2051/1-1.
\end{acknowledgments}

\appendix
\section{Cumulants}
\label{app:Cumulants}	
The cumulants can be represented in terms of products of moments as \cite{mendelIEEE1991,gardinerBOOK2009}
	\begin{eqnarray}
	C_2(x,y) & = & \langle yx \rangle -\langle y \rangle \langle x \rangle, \label{eq:C2}  \\
	C_3(x,y,z) & = & \langle zyx \rangle - \langle yx \rangle\langle z \rangle \nonumber\\
	&-& \langle zx \rangle\langle y \rangle
	- \langle zy \rangle\langle x \rangle + 2 \langle z \rangle\langle y \rangle\langle x \rangle, \label{eq:C3} \\
C_4(x,y,z,w) & = & \langle wzyx \rangle - \langle wzy \rangle\langle x \rangle - \langle wyx \rangle\langle z \rangle\nonumber \\
&-&\langle wzx \rangle\langle y \rangle-\langle zyx \rangle\langle w \rangle -\langle wz \rangle\langle yx \rangle \nonumber \\ \nonumber
&-&  \langle wy \rangle\langle zx \rangle-\langle wx \rangle\langle zy \rangle +2\langle yx \rangle\langle w \rangle \langle z \rangle \\ \nonumber
&+&2\langle zx \rangle\langle w \rangle \langle y \rangle+2\langle wx \rangle\langle y \rangle \langle z
\rangle \\ \nonumber
&+& 2\langle wy \rangle\langle z \rangle \langle x \rangle+2\langle zy \rangle\langle w \rangle \langle x \rangle \\ 
&+& 2\langle wz \rangle\langle y \rangle \langle x \rangle - 
 6\langle x \rangle\langle y \rangle\langle z \rangle\langle w \rangle. \label{eq:C4}
\end{eqnarray}
\section{Polyspectra and their estimation from time traces}
\label{app:polyspectra}
Starting from the  auto-correlation function of the detector output $z(t)$ 
\begin{eqnarray}
  a(\tau) & = &  C_2(z(t),z(t+\tau)) \nonumber \\
  & = & \left\langle z(t) z(t+\tau) \right\rangle_t - \left\langle z(t) \right\rangle_t\left\langle z(t+\tau)\right\rangle_t, 
\end{eqnarray}
where $\left\langle ... \right\rangle_t$ relates to the ideal infinite time average with respect to $t$,
 the power spectrum   
\begin{eqnarray}
S_z^{(2)}(\omega) &=& \int_{-\infty}^{\infty}a(\tau)e^{i\omega\tau}\textrm{d}\tau
\end{eqnarray}
 can be defined.  
Alternatively, the power spectrum
 can be expressed via the Fourier transform of the detector output  $z(\omega) = \int_{-\infty}^{\infty}z(t)e^{i\omega t}\textrm{d}t$
 as
 \begin{eqnarray}
   2\pi \delta(\omega+\omega')S_z^{(2)}(\omega) &=& C_2(z(\omega), z(\omega')).
\end{eqnarray} 
Brillinger generalized this expression to define polyspectra of order $n$  (Ref. \onlinecite{Brillinger1965})
\begin{eqnarray}
2\pi \delta(\omega_1+...+\omega_n)&S_z^{(n)}&(\omega_1,...,\omega_{n-1})\nonumber\\ 
&=& C_n(z(\omega_1),..., z(\omega_n)). \label{eq:defPolyspectra}
\end{eqnarray}
 \begin{figure}[t]
	\begin{center}
		\includegraphics[width=6.5cm]{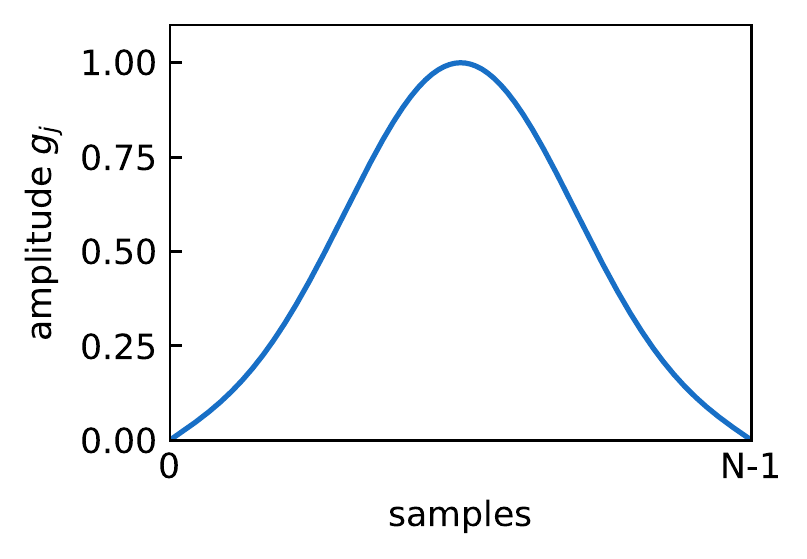}\\[3mm]
		\caption[Window function]{Parameter dependent approximate confined Gaussian window function $g_j$ for parameter $s = 0.14 T$ \cite{starosielecSP2014}.}
		\label{soft}
	\end{center}
\end{figure}
Above, the bispectrum
\begin{eqnarray}
  S_z^{(3)}(\omega_1,\omega_2),
\end{eqnarray}
and a cut through the trispectrum $S_z^{(4)}$ 
\begin{eqnarray}
S_z^{(4)}(\omega_1,\omega_2) = S_z^{(4)}(\omega_1,-\omega_1,\omega_2).
\end{eqnarray}
 are used for characterizing experimental time traces and comparison with quantum polyspectra.
 Polyspectra are estimated from experimental time traces in the following way:
 The detector output $z(t)$ is discretized and divided into time frames leading to arrays $z^{(n)}$ of length $N$ with $0 \le j < N$
 \begin{equation}
   z^{(n)}_j = z(j T/N + n T),
 \end{equation}  
 where $T$ is the temporal length of the time frames. 
 The coefficient $a^{(n)}_k$ of the discrete Fourier transformation (fast Fourier transformation can be used for evaluation) are obtained after applying a window function  $g_j$ to $z_j$
 \begin{equation}
 a^{(n)}_k = \frac{T}{N} \sum_{j=0}^{N-1} g_j z^{(n)}_j e^{2 \pi i j k / N}.
 \end{equation}
 Window functions are routinely used in signal processing for improving the spectral resolution \cite{harrisProc1978}.
Here we appy the approximated confined Gaussian window with window parameter $s = 0.14 T$ (Figure \ref{soft}) for its optimal RMS time-bandwidth product \cite{starosielecSP2014}.

The ideal polyspectra are then approximately (finite spectral resolution) given by \cite{Schefczik2020}
\begin{eqnarray}
S^{(2)}_z(\omega_k) &\approx& \frac{N C_2(a_k, a_k^*)}{T \sum_{j = 0}^{N-1} g_j^2} \\
S^{(3)}_z(\omega_k, \omega_l) &\approx& \frac{N C_3(a_k, a_l, a_{k+l}^*)}{T \sum_{j = 0}^{N-1} g_j^3} \\
S^{(4)}_z(\omega_k, \omega_l) &\approx& \frac{N C_4(a_k, a_k^*, a_l, a_l^*)}{T \sum_{j = 0}^{N-1} g_j^4},
\end{eqnarray}
where $\omega_k = 2 \pi k / T$ for $k< N/2$ and $\omega_k = 2 \pi (k - N)/T$ for $k \ge N/2$. 
The cumulants $C_2$, $C_3$, and $C_4$ (see App. \ref{app:Cumulants}) are estimated from so-called cumulant estimators \cite{schefczikARXIV2019}
\begin{eqnarray}
c_2(x,y) =&& \frac{m}{m-1}\left( \overline{xy} - \overline{x} \,\overline{y} \right) \\
c_3(x,y,z) =&& \frac{m^2}{(m-1)(m-2)}\nonumber \\&&\times\left( \overline{xyz}- \overline{xy}\, \overline{z} - \overline{xz}\, \overline{y} \right. \nonumber \\
 &&- \left. \overline{yz}\, \overline{x} + 2 \overline{x}\,\overline{y}\, \overline{z} \right) 
\end{eqnarray}
\begin{eqnarray}
 c_4(x,y,z,w) =&& \frac{m^2}{(m-1)(m-2)(m-3)} \nonumber \\
&& \nonumber \hspace{-1cm} \times \Big[(m+1) \overline{xyzw} - (m+1) \left(\overline{xyz} \ \overline{w} + \textrm{ 3 o.p.}\right) \Big. \\ 
&&\nonumber \Big. -  (m-1) \left(\overline{xy} \ \overline{zw} + \textrm{2 o.p.} \right)  \Big. \\
&& \Big.+ 2 m \left(\overline{xy} \ \overline{z} \ \overline{w} +\textrm{5 o.p.}\right) \Big. \nonumber \\
&&- \Big. 6m \overline{x} \ \overline{y} \ \overline{z} \ \overline{w} \Big.],
\end{eqnarray}
where o.p. means "other permutations". The overline  $\overline{( ...)}$ denotes an average of $m$ samples. 
Their structure is similar to that of the cumulants apart from $m$-dependent prefactors [compare Eqs. (\ref{eq:C2})-(\ref{eq:C4})]. 
The estimators have the property
$\langle c_j \rangle = C_j$ for finite $m$ (unbiased estimators) and $c_j \rightarrow C_j$ for $m \rightarrow \infty$ (consistency). The estimators $c_2$, $c_3$, and $c_4$ are multivariate versions of the well-known k-statistics \cite{FisherPLMS1928,kendallBOOK1943,mccullaghBOOK2018,cookBIOMETRIKA1951}. The estimator $c_2(x,x)$ is identical with a frequently used 
estimator for the variance of $x$. It exhibits the typical prefactor $m/(m-1)$ which is sometimes called the Bessel-correction \cite{kenneyBOOK1951}. 
 
\section{Quantum polyspectra}
\label{app:QuantumPolyspectra}
The polyspectra of detector output $z(t)$  of the continuously monitored quantum system in the steady state follow from the SME
without any approximations as
	\begin{widetext}
	\begin{eqnarray}	
	S_z^{(2)}(\omega) &=& \beta^4 ( {\rm Tr}[{\cal A}'{\cal G}'(\omega){\cal A}'\rho_0] +  {\rm Tr}[{\cal A}'{\cal G}'(-\omega){\cal A}'\rho_0] ) + \beta^2/4 \label{eq:S2}\\	S_z^{\rm (3)}(\omega_1,\omega_2,\omega_3 = -\omega_1-\omega_2)  &= & 
	\beta^6 \sum_{\text{prm. $\omega_1,\omega_2,\omega_3$}}
	\hspace{-4mm} {\rm Tr}[{\cal A}'{\cal G}'(\omega_3){\cal A}'{\cal G}'(\omega_3 + \omega_2){\cal A}'\rho_0]. \label{eq:S3}  \\
	S_z^{\rm (4)}(\omega_1,\omega_2,\omega_3,\omega_4 = -\omega_1-\omega_2-\omega_3) & = & \beta^8  \sum_{\text{prm. $\omega_1,\omega_2,\omega_3,\omega_4$}}
	\hspace{-8mm} \left[ {\rm Tr}[{\cal A}'{\cal G}'(\omega_4){\cal A}' {\cal G}'(\omega_3 + \omega_4){\cal A}'{\cal G}'(\omega_2 + \omega_3 + \omega_4){\cal A}'\rho_0] \right.  \label{eq:S4} \\  \nonumber
	&-& \frac{1}{2 \pi}\int{\rm Tr}[{\cal A}'{\cal G}'(\omega_4) {\cal G}'(\omega_3 + \omega_4 - \omega){\cal A}'\rho_0]{\rm Tr}[{\cal A}'{\cal G}'(\omega) {\cal G}'(\omega_2 +\omega_3 + \omega_4){\cal A}'\rho_0]\textrm{d}\omega \\ \nonumber
	&-& \left. \frac{1}{2 \pi}\int{\rm Tr}[{\cal A}'{\cal G}'(\omega_4) {\cal G}'(\omega_2 +\omega_3 + \omega_4){\cal G}'(\omega_3 + \omega_4 - \omega){\cal A}'\rho_0]{\rm Tr}[{\cal A}'{\cal G}'(\omega) {\cal A}'\rho_0]\textrm{d}\omega\right]. 
	\end{eqnarray}
Their derivation via multi-time cumulants of $z(t)$ and an efficient method for their numerical evaluation are given in Refs. \onlinecite{hagelePRB2018,hagelePRB2020E}.
All spectra are free from delta-function contributions because the time-dependent ${\cal G}'(\tau)$ decays exponentially to zero for increasing $\tau$ as long as existence of a  steady state $\rho_0 = {\cal G}(\tau)\rho$ for $\tau \rightarrow \infty$ can be assumed.

\section{The fourth-order trispectrum of the SQD}
\label{app_s4}
Analytical expression for the trispectrum of the quantum dot system (neglecting spin-dependent dynamics):
\begin{eqnarray}
\label{eq:s4xx}
S^{(4)}_z(\omega_1, \omega_2) &=& 4 \gamma _{\text{in}} \gamma _{\text{out}} \left(\gamma _{\text{in}}^2 \left(\left(\gamma _{\text{in}}+\gamma _{\text{out}}\right){}^2+\omega _1^2\right) \left(\left(\gamma _{\text{in}}+\gamma _{\text{out}}\right){}^2+\omega _2^2\right) \left(3 \left(\gamma _{\text{in}}+\gamma _{\text{out}}\right){}^2 \left(2 \left(\gamma _{\text{in}}+\gamma _{\text{out}}\right){}^2 \right.\right.\right.\nonumber\\
&+& \left.\left. \left. \omega _1^2+\omega _2^2\right)+\left(\omega _1^2-\omega _2^2\right){}^2\right)-2 \gamma _{\text{in}} \gamma _{\text{out}} \left(\left(\gamma _{\text{in}}+\gamma _{\text{out}}\right){}^2+\omega _1^2\right) \left(\left(\gamma _{\text{in}}+\gamma _{\text{out}}\right){}^2+\omega _2^2\right)\right.\nonumber\\
&\times& \left.\left(3 \left(\gamma _{\text{in}}+\gamma _{\text{out}}\right){}^2 \left(2 \left(\gamma _{\text{in}}+\gamma _{\text{out}}\right){}^2+\omega _1^2+\omega _2^2\right)+\left(\omega _1^2-\omega _2^2\right){}^2\right)+2 \gamma _{\text{in}} \gamma _{\text{out}} \left(\left(\gamma _{\text{in}}+\gamma _{\text{out}}\right){}^2+\left(\omega _1-\omega _2\right){}^2\right) \right.\nonumber\\
&\times&\left. \left(\left(\gamma _{\text{in}}+\gamma _{\text{out}}\right){}^2+\left(\omega _1+\omega _2\right){}^2\right) \left(\omega _1^2 \omega _2^2-\left(\gamma _{\text{in}}+\gamma _{\text{out}}\right){}^2 \left(3 \left(\gamma _{\text{in}}+\gamma _{\text{out}}\right){}^2+\omega _1^2+\omega _2^2\right)\right)\right.\nonumber\\
&+&\left. \gamma _{\text{out}}^2 \left(\left(\gamma _{\text{in}}+\gamma _{\text{out}}\right){}^2+\omega _1^2\right) \left(\left(\gamma _{\text{in}}+\gamma _{\text{out}}\right){}^2+\omega _2^2\right) \left(3 \left(\gamma _{\text{in}}+\gamma _{\text{out}}\right){}^2 \right.\right.\nonumber\\
&\times&\left.\left.\left(2 \left(\gamma _{\text{in}}+\gamma _{\text{out}}\right){}^2+\omega _1^2+\omega _2^2\right)+\left(\omega _1^2-\omega _2^2\right){}^2\right)\right)\nonumber\\
&/&\left(\gamma _{\text{in}}+\gamma _{\text{out}}\right){}^3 \left(\left(\gamma _{\text{in}}+\gamma _{\text{out}}\right){}^2+\omega _1^2\right){}^2 \left(\left(\gamma _{\text{in}}+\gamma _{\text{out}}\right){}^2+\left(\omega _1-\omega _2\right){}^2\right)\nonumber\\
&/& \left(\left(\gamma _{\text{in}}+\gamma _{\text{out}}\right){}^2+\omega _2^2\right){}^2 \left(\left(\gamma _{\text{in}}+\gamma _{\text{out}}\right){}^2+\left(\omega _1+\omega _2\right){}^2\right)
\end{eqnarray}
\end{widetext}

\end{document}